\documentclass[%
reprint,
showpacs,preprintnumbers,
nofootinbib,
amsmath,amssymb,
prb,
]{revtex4-1}

\usepackage{mathptmx}
\usepackage{epstopdf}
\usepackage{graphicx}
\usepackage{dcolumn}
\usepackage{bm}
\usepackage{hyperref}
\usepackage{ulem}
\usepackage{color}

\newcommand{\mean}[1]{\langle#1\rangle}

\newcommand{\etc}{{\it etc}}
\newcommand{\hence}{\ \Rightarrow\ }
\newcommand{\partd}[2]{\displaystyle\frac{\partial#1}{\partial#2}}
\newcommand{\svector}[2]{\left(\begin{array}{c}#1 \\ #2 \end{array}\right)}
\newcommand{\smatrix}[4]{\left(\begin{array}{cc}#1 & #2 \\ #3 & #4 \end{array}\right)}

\graphicspath{ {./Figures//} }

\begin{document}

\title{Improving the phase super-sensitivity of squeezing-assisted interferometers by squeeze factor unbalancing}

\author{Mathieu Manceau$^{1}$, Farid Khalili$^{3}$ and Maria Chekhova$^{1,2,3}$}

\address{$^1$Max-Planck-Institute for the Science of Light, Erlangen, Germany\\
$^2$University of Erlangen-N\"urnberg, Staudtstrasse 7/B2, 91058 Erlangen, Germany\\
$^3$Faculty of Physics, M V Lomonosov Moscow State University, Moscow, Russia}

\begin{abstract}
The sensitivity properties of  an SU(1,1) interferometer made of two cascaded parametric amplifiers, as well as of an ordinary SU(2)
interferometer preceded by a squeezer and followed by an anti-squeezer, are theoretically investigated. Several possible experimental
configurations are considered, such as the absence or presence of a seed beam, direct or homodyne detection scheme. In all cases we
formulate the optimal conditions to  achieve phase super-sensitivity, meaning a sensitivity overcoming the shot-noise limit. We show that
for a given gain of the first parametric amplifier, unbalancing the interferometer by increasing the gain of the second amplifier improves
the interferometer properties. In particular, a broader super-sensitivity phase range  and a better overall sensitivity can be achieved by
gain unbalancing.
\end{abstract}


\maketitle

\section{Introduction}\label{sec:Intro}

Estimating the phase of light is one of the most important tasks in optics. It is the basic tool in many fields, from spectroscopy to the
gravitational wave detection~\cite{PRL_116_061102_2016}. A straightforward way to assess an optical phase shift is to use an interferometer.
The phase sensitivity of an interferometer has fundamental bounds, given by the quantum
properties of light.  In particular, in the most
basic case of a coherent quantum state, the phase sensitivity is
\begin{equation}\label{SNL}
  \Delta\phi_{\rm SNL} = \frac{1}{2\sqrt{N}} \,,
\end{equation}
where $N$ is the mean photon number \footnote{This value differs from the one frequently found in the literature by the factor $1/2$. The 
same is true for the Heisenberg limit \eqref{HL} and other equations for the phase sensitivity in two-arm schemes. This factor arises 
because we define the relative phase shift in a two-arm scheme as $2\phi$. This definition provides more consistent equations for the two- 
and single-arm cases.}. This restriction is known as the shot noise limit (SNL), due to the Poissonian photon statistics of the coherent 
quantum state.

It is known that the SNL can be overcome by using more advanced sources of light. In fact, the ultimate phase sensitivity achievable
according to quantum mechanics is much better than the SNL --- it is given by the Heisenberg limit (HL),
\begin{equation}\label{HL}
  \Delta\phi_{\rm HL} \sim \frac{1}{2N} \,,
\end{equation}
which is named so because it could be considered as a consequence of the Heisenberg uncertainty relation for the number of quanta and the
phase~\cite{Heitler1954},
\begin{equation}\label{dNdPhi}
  \Delta N\Delta\phi \ge \frac{1}{2} \,.
\end{equation}
$N$ is non-negative, therefore, $\Delta N\le N$, which gives \eqref{HL}.\footnote{Strictly speaking, this ``proof'' is incorrect, because
(i) a ``well-behaved'' phase operator can not be defined \cite{Carruthers_RMP_40_411_1968} and (ii) there exist probability distributions
with $\Delta N\gg N$. Nevertheless, substitutes of the phase operator are proposed which give the same result \eqref{dNdPhi} provided that
$\Delta\phi\ll1$ \cite{Susskind_Physics_1_49_1964, Carruthers_RMP_40_411_1968, Popov_VLU_22_7_1973, Pegg_PRA_39_1665_1989}.}

The ultimate limit for the phase sensitivity can be obtained using the Rao-Cramer approach \cite{HelstromBook}. It has the form of the
uncertainty relation \eqref{dNdPhi}, but with a different meaning of $\Delta\phi$ --- now it is a phase shift imposed by some external
agent, and not the variance of the (non-existing) phase operator. This limit assumes that some optimal measurement procedure on the outgoing
light is used. Whether this procedure can be implemented in practice is another issue. However, it is known that at least Gaussian quantum
states of light, if fed into certain interferometers, can saturate the Rao-Cramer bound in the ideal lossless case
\cite{Caves1981,Yurke_PRA_A_33_4033_1986,Braunstein1994,pezze2008}.

\begin{figure*}
  \centering\includegraphics[width=0.7\textwidth]{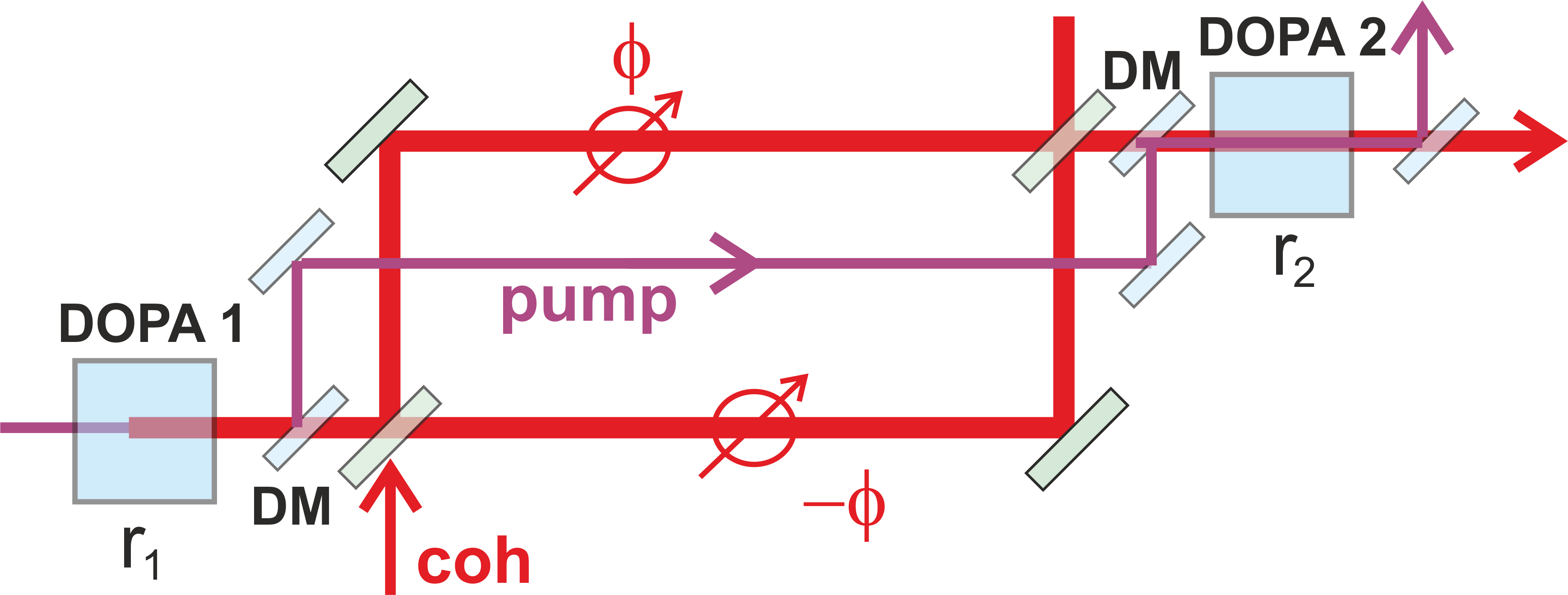}
  \caption{A linear interferometer preceded by a squeezer (degenerate optical parametric amplifier, DOPA1, with the
squeeze factor $r_1$) and followed by an anti-squeezer (DOPA2), with the squeeze factor $r_2$. The pump path can be
arranged by using dichroic mirrors DM.}
\label{fig:linear}
\end{figure*}

It was shown by direct calculations that the sensitivity \eqref{HL} can be achieved using exotic quantum states like
Pegg-Barnett~\cite{Holland1993} or NOON~\cite{lee2002quantum} ones.  However, generation of these states for $N\gg1$ is impossible with the
current state-of-the-art technologies. Another, much more realistic class of quantum states, namely squeezed states, was explored in the
pioneering work by C.\,Caves \cite{Caves1981}. In the reasonable case of not very strong squeezing, $e^{2r}\ll N$, where $r$ is the squeeze
factor, the phase sensitivity corresponds to the ``improved'' SNL:
\begin{equation}\label{dphi_sqz}
\Delta\phi_{\rm sqz} =
\frac{e^{-r}}{2\sqrt{N}} \,.
\end{equation}
In the (hypothetic) case of very strong squeezing, $e^{2r}\sim N$, $\Delta\phi_{\rm sqz}$ asymptotically approaches the Heisenberg limit
\eqref{HL}.

Recently, reduction of the noise below the SNL by means of squeezed-states injection was demonstrated in kilometers-scale interferometers of
laser gravitational-wave detectors GEO-600 \cite{Nature_2011} (which continues to routinely operate in this regime \cite{GEOsite}) and LIGO
\cite{Nature_2013}. A $10$\,dB squeezed light source was used in these experiments. However, the achieved sensitivity gain was much smaller
(about $3\,{\rm dB}$ and $2\,{\rm dB}$, respectively). The reason for such modest effect of squeezing on the phase sensitivity is the
fragility of squeezed light to optical losses, both inside the interferometer (the {\it internal losses}) and in the output optical path,
including the photodetectors' quantum inefficiency (the {\it external losses}). It should be noted that in the state-of-the-art
interferometers and, in particular, in laser gravitational-wave detectors, it is the external losses that apply most severe limitations on
the sensitivity  (see the loss budget analysis in \cite{Nature_2011, Nature_2013}).

In the same paper~\cite{Caves1981}, C.\,Caves proposed to amplify the signal, as well as the noise component that is in phase with the
signal, by means of a second (anti)squeezer located in the output path of the interferometer. In Fig.\,\ref{fig:linear}, implementation of
this idea for a Mach-Zehnder interferometer is shown. This additional squeezer does not affect the signal-to-noise ratio (SNR) associated
with the internal losses and the  corresponding noise, because both the signal and this noise are equally amplified or de-amplified. At the
same time, amplification of the signal by the output anti-squeezer suppresses the influence of the external losses on the phase sensitivity.
It is important to note that the decrease of $\Delta\phi$ is not accompanied by an increase of $\Delta N$, see Eq.\,\eqref{dNdPhi}, because
the optical quantum state inside the interferometer is obviously not affected by the output anti-squeezer. This means that the phase
measurement indeed becomes more efficient, that is, closer to the Rao-Cramer bound.

In 1986, an ingenious scheme, the so-called {\it SU(1,1) interferometer}, was proposed by Yurke et al.~\cite{Yurke_PRA_A_33_4033_1986}. It
can be viewed as a further development of the idea of Fig.\,\ref{fig:linear}, with the beam-splitters ``fused'' together with the
corresponding squeezers.  Two versions of this scheme were proposed, based on the degenerate (DOPA) and non-degenerate (NOPA) optical
parametric amplifiers, shown in  Fig.~\ref{fig:SU(1,1)}a,b, respectively. Here, light generated or amplified in the first OPA is amplified
or deamplified in the second one, depending on the phase shifts $\phi_{p,s,i}$ introduced in the pump, signal, and idler channels,
respectively. The term SU(1,1) comes from the fact that this interferometer performs Bogolyubov (related to the SU(1,1) group)
transformations of the optical fields (note that the SU(1,1) interferometer is a special case of a broader class of nonlinear
interferometers~\cite{Chekhova2016}). Correspondingly, for the ordinary linear interferometers, performing an SU(2) transformation, the term
{\it SU(2) interferometer} was coined in \cite{Yurke_PRA_A_33_4033_1986}; we adopt this terminology here.

\begin{figure*}
\centering\includegraphics[width=0.7\textwidth]{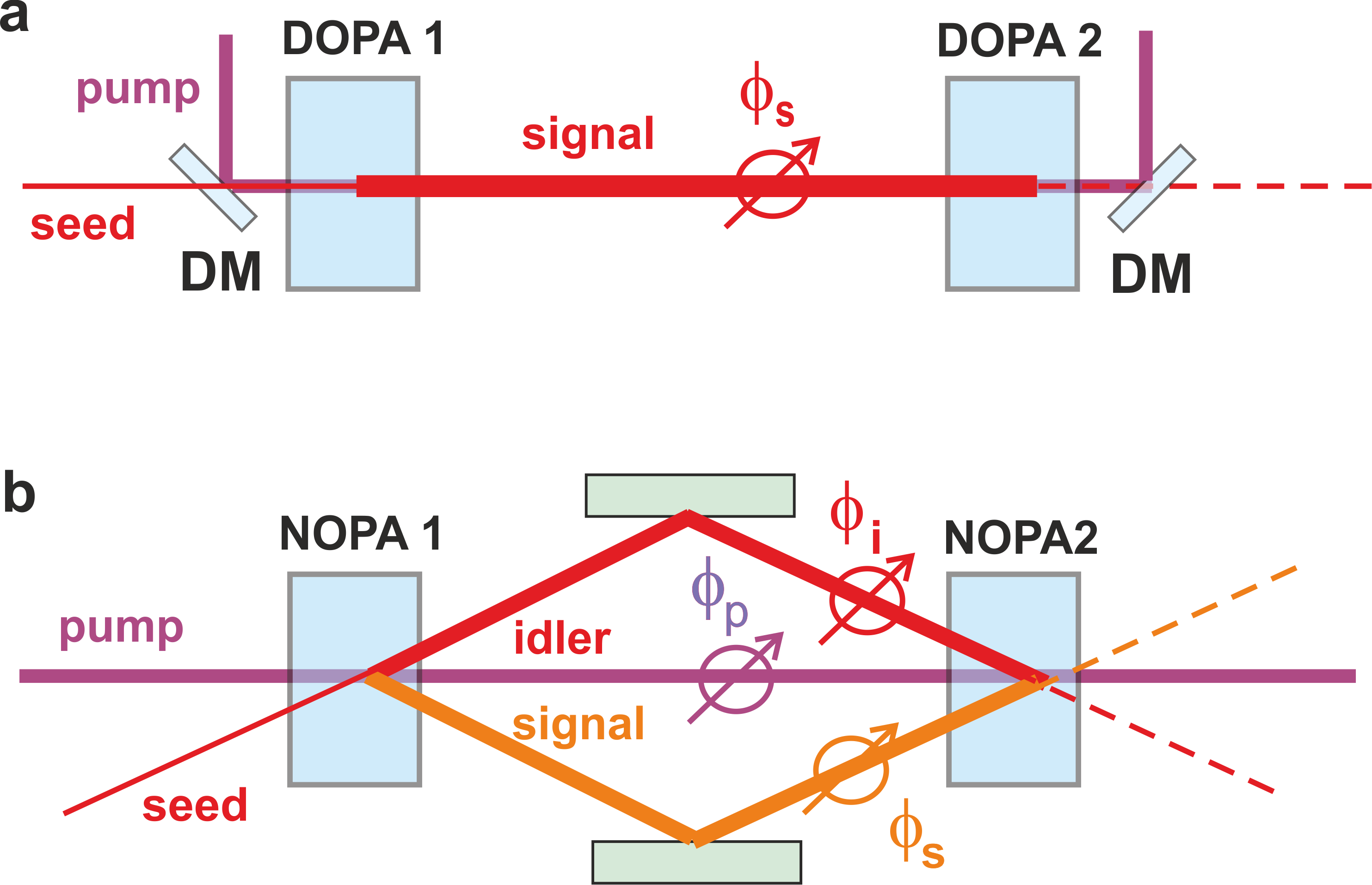}
\caption{Degenerate (a) and non-degenerate (b) SU(1,1) interferometers.}
\label{fig:SU(1,1)}
\end{figure*}

Later, Plick et al. showed~\cite{Plick_NJP_12_083014_2010} that the performance of an SU(1,1) interferometer can be
improved by seeding it with coherent light.

The influence of the optical losses on the performance of this scheme was calculated by Marino et al.~
\cite{Marino_PRA_86_023844_2012} who explicitly showed it to be highly immune to external losses. This result is well
expected, taking into account that the SU(1,1) interferometer uses the same squeezing/antisqueezing principle as the
scheme of C.~Caves.

Recently, an SU(1,1) interferometer has been implemented using two cascaded four-wave mixers (FWM), and an enhancement
in the fringe intensity compared to a linear interferometer  without squeezing was
demonstrated~\cite{Jing_APL_99_011110_2011, Kong_APL_102_011130_2013}. In Ref.~\cite{Hudelist_NComms_5_3049_2014}, about
4\,dB enhancement of the phase sensitivity in an SU(1,1) interferometer was achieved. Phase locking aimed at working at
the optimum sensitivity point of such an interferometer was also demonstrated \cite{Wang_APL_107_121106_2015}.

Evidently, the signs of the squeeze factors $r_1,\,r_2$, given by the parametric gain values of the first and the second
OPAs, both in the linear (Fig.\,\ref{fig:linear}) and non-linear (Fig.\,\ref{fig:SU(1,1)}) interferometers, have to be
opposite, $r_1r_2<0$. Their absolute values could be equal (the balanced case, $|r_1|=|r_2|$) or different (the
unbalanced case, $|r_1|\ne|r_2|$). Typically in the previous works, starting from the initial paper~\cite{Caves1981},
the balanced case was considered theoretically and implemented experimentally. This is a necessary requirement in the
case of FWM, as the mode structure of an FWM OPA considerably depends on the gain~\cite{Liu_OE_24_1096_2016}. However,
it does not have to be the case in the other realizations of the parametric amplifiers. In particular, it was shown
in~\cite{Sharapova_PRA_91_043816_2015} that the shapes of spatial and temporal modes of the parametric amplifiers based
on high-gain parametric down-conversion~\cite{Iskhakov_OL_37_1919_2012, Perez_OL_39_2403_2014} can be
considered as gain-independent. The unbalanced regime of an SU(1,1) interferometer was considered in
Ref.~\cite{Li_NJP_16_073020_2014}, with the conclusion that the best regime is the balanced or close to the balanced
one. However, this conclusion was based on a non-optimal detection procedure, with only one of the two output beams
measured.

In this paper, we consider the unbalanced regime in detail and show that if a proper detection procedure is used, then indeed the sensitivity for both 
SU(2) and SU(1,1) interferometers increases with the output squeezing strength. We focus here on  ``practical'' schemes involving Gaussian (coherent 
or squeezed) states of light and available detection methods. The general introduction into the field can be found {\it e.g.} in the review article 
\cite{Demkowicz_PIO_60-345_2015}.

An important feature of the phase estimation is that it is essentially a relative measurement and assumes the existence of some phase 
reference point in the form of {\it e.g.} the local oscillator phase or the optical pump phase. (Probably the only exception is a 
non-squeezed two-arms linear interferometer with the direct photocounting at the output(s).) This phase reference beam could be viewed as a 
problem in practical realizations of phase-supersensitive schemes. Indeed, the photon number $N$ entering the expressions for SNL and HL is 
actually limited by various undesirable effects of the optical power incident on the probe objects, from technical ones like heating to 
truly fundamental, like the quantum radiation pressure fluctuations. However, these reference beams do not have to interact directly with 
the probe objects. For this reason, we assume that the optical power in these beams could be as high as necessary to make the additional 
measurement error imposed by the phase uncertainties of these beams negligibly small.

In the next section, we analyze the unbalanced regime of an SU(2) interferometer preceded by a squeezer and followed by an anti-squeezer. Homodyne 
detection at the output of the interferometer is considered. Then in Section \ref{sec:SU(1,1)Seeded} we consider an unbalanced SU(1,1) interferometer 
with coherent seeding and the optimal homodyne detection at the output. Beyond the increased sensitivity with the output squeezing strength, we show 
that both types of interferometer share similar equations defining their phase sensitivity properties.

In Secs.\,\ref{sec:SU(1,1)SeededDirect} and \,\ref{sec:SU(1,1)Unseeded} we present the case of an SU(1,1)
interferometer with the (less optimal) direct detection scheme, with and without coherent seeding. Here
as well, the phase sensitivity is shown to increase with the output squeezing strength.

Finally, a comparison between different interferometers and detection schemes is provided in Sec.\ref{sec:Comparison},
the effects of the first squeezer strength is considered.

In the Appendices, we provide detailed calculations for each case addressed in this paper. We also show that a
non-degenerate SU(1,1) interferometer can be treated as two independent degenerate ones. For this reason, both in
Section~\ref{sec:SU(1,1)Seeded} and in  Section~\ref{sec:SU(1,1)Unseeded} we consider a degenerate SU(1,1)
interferometer based on two DOPAs.

\section{Linear interferometer with squeezed input}\label{sec:Linear}

In this section we consider the scheme shown in Fig.~\ref{fig:linear}, based on an ordinary Mach-Zehnder interferometer.
Squeezed vacuum from DOPA1 is injected into one of the input ports; the second port is fed with coherent light with the
normalized amplitude $\alpha$. If the interferometer is perfectly symmetric, then the coherent light leaves through
one of the output ports (the ``bright'' one), and the squeezed vacuum, through the other (``dark'') port. If the light
in the interferometer arms experiences some antisymmetric phase shifts $\phi_1=\phi$ and $\phi_2=-\phi$, then a fraction of the coherent
light is redirected to the dark port, displacing the output squeezed state by $\alpha\phi$. Then the dark port output
is amplified by DOPA2 and then measured. In the original work \cite{Caves1981}, direct measurement of the number of quanta was considered. 
Here, for the sake of completeness, we consider the measurement of the quadrature containing the phase information by means of a homodyne 
detector.

In the more general case of arbitrary phase shifts $\phi_1$ and $\phi_2$, they can be decomposed into the symmetric ($\psi$) and 
antisymmetric ($\phi$) components. It can be shown that in this case the light emerging from the dark port still carries out information 
about $\phi$, while the phase of the light in the bright port depends on $\psi$. Therefore, in principle, the symmetric phase shift $\psi$ 
also could be measured in this scheme by using the second homodyne detector in the bright port.

Note that in the particular case of $\phi_1=2\phi$ and $\phi_2=0$, both output ports carry the information about the same phase $\phi$. 
In this case, the bright port homodyne detector allows one to retrieve additional information about $\phi$. Therefore, as it was shown in 
\cite{Jarzina_PRA_85_011801_2012}, this case, in principle, provides better phase sensitivity than the antisymmetric one with the same phase 
difference between the arms, $\phi_1=\phi$ and $\phi_2=-\phi$. However, in the real-world interferometers (laser gravitational-wave 
detectors can be mentioned as the most conspicuous example), the bright port is contaminated by the laser technical noises and therefore 
this possibility can not be considered as a practical one. Here we limit ourselves to the antisymmetric case only.

Following the initial proposals~\cite{Caves1981, Yurke_PRA_A_33_4033_1986}, both here and in the following section,
devoted to the SU(1,1) case, we assume that both squeezers are in phase with the coherent light. In this case, it
is natural to set the corresponding phases equal to zero. In particular, this means that $\alpha$ is real. We
assume also the following signs of the squeeze factors: $r_1>0$, $r_2<0$.

We show in \ref{app:linear} that the phase sensitivity in this case is
\begin{equation}\label{lin:dPhi}
  (\Delta\phi)^2 = (\Delta\phi_{\rm min})^2 + K\tan^2\phi \,,
\end{equation}
where
\begin{equation}\label{lin:dPhi_min}
  \Delta\phi_{\rm min} = \frac{1}{2\alpha}
    \sqrt{e^{-2r_1} + \frac{1-\mu}{\mu} + \frac{1-\eta}{\mu\eta}\, e^{-2|r_2|}},
\end{equation}
is the best sensitivity achieved at $\phi=0$, and
\begin{equation}\label{lin:K}
  K = \frac{1}{4\alpha^2}
    \left(\frac{1}{\mu} + \frac{1-\eta}{\mu\eta}\, e^{-2|r_2|}\right)
\end{equation}
is the factor defining the sensitivity deterioration with the increase of $\phi$. In Eqs.~(\ref{lin:dPhi},\ref{lin:K}), $\mu$ and $\eta$ are the 
quantum efficiencies corresponding to the internal and the external losses, respectively. One can easily see that at  $|r_2|\to\infty$, the terms in 
$\Delta\phi_{\rm min}$ and $K$ associated with the external losses vanish.

Note that despite the different detection procedures, we obtained virtually the same result as in \cite{Caves1981}, up to the notation and taking into 
account that the condition $r_1=|r_2|$ was assumed in that paper. The characteristic dependence of the phase sensitivity on the losses in 
Eqs.\,(\ref{lin:dPhi_min}, \ref{lin:K}), as well as in other similar equations below in this paper, actually corresponds to the fundamental bound on 
lossy interferometry predicted in \cite{Knysh_PRA_83_021804_2011, Escher_NPhys_7_406_2011, Demkowicz_NComm_3_1063_2012} and observed experimentally in 
\cite{Demkowicz_PRA_88_041802_2013}.

We consider first the optimization of $\Delta\phi_{\rm min}$ with respect to the input squeezing strength $r_1$ for a
given mean number of photons used for the measurement,
\begin{equation}\label{lin:N}
  N = \alpha^2 + \sinh^2r_1,
\end{equation}
in the ideal lossless case $\mu=\eta=1$. In this case, the minimum of \eqref{lin:dPhi_min} in $r_1$ occurs for
\begin{equation}\label{opt_r_1}
  e^{2r_1} = 2N+1 \,,
\end{equation}
and is equal to
\begin{equation}\label{min_dphi}
  (\Delta\phi_{\rm min})^2 = \frac{1}{4N(N+1)} \,.
\end{equation}
This means that the setup shown in Fig.~\ref{fig:linear} could reach the HL in the ideal lossless scenario.

In addition to $\Delta\phi_{\rm min}$, another important figure of merit is the phase range $\Delta$ where the
sensitivity exceeds the SNL. We will further call it the supersensitive phase range. It follows from \eqref{lin:dPhi}
that it is equal to
\begin{equation}\label{Delta}
  \Delta = 2\arctan\sqrt{\frac{(\Delta\phi_{\rm SNL})^2 - (\Delta\phi_{\rm min})^2}{K}}\,.
\end{equation}
For a high-precision measurement, $\Delta$ approaches a simple asymptotic value. Suppose that
\begin{equation}\label{rw_cond_1}
  (\Delta\phi_{\rm min})^2 \ll (\Delta\phi_{\rm SNL})^2 \,.
\end{equation}
The necessary conditions for this are
\begin{equation}
  1-\mu \ll 1 \,, \quad (1-\eta) e^{-2|r_2|} \ll 1 \,.
\end{equation}
Note also that in the reasonable real-world scenarios,
\begin{equation}\label{rw_cond_2}
  \alpha^2\gg\sinh^2r_1 \hence N\approx\alpha^2 \,.
\end{equation}
Below we assume this condition for all relevant cases. In particular, it follows from \eqref{rw_cond_2} that $K\approx(\Delta\phi_{\rm SNL})^2$ and
\begin{equation}
  \Delta \approx \frac{\pi}{2}
\end{equation}

\begin{figure}
  \centering \includegraphics[scale=0.5]{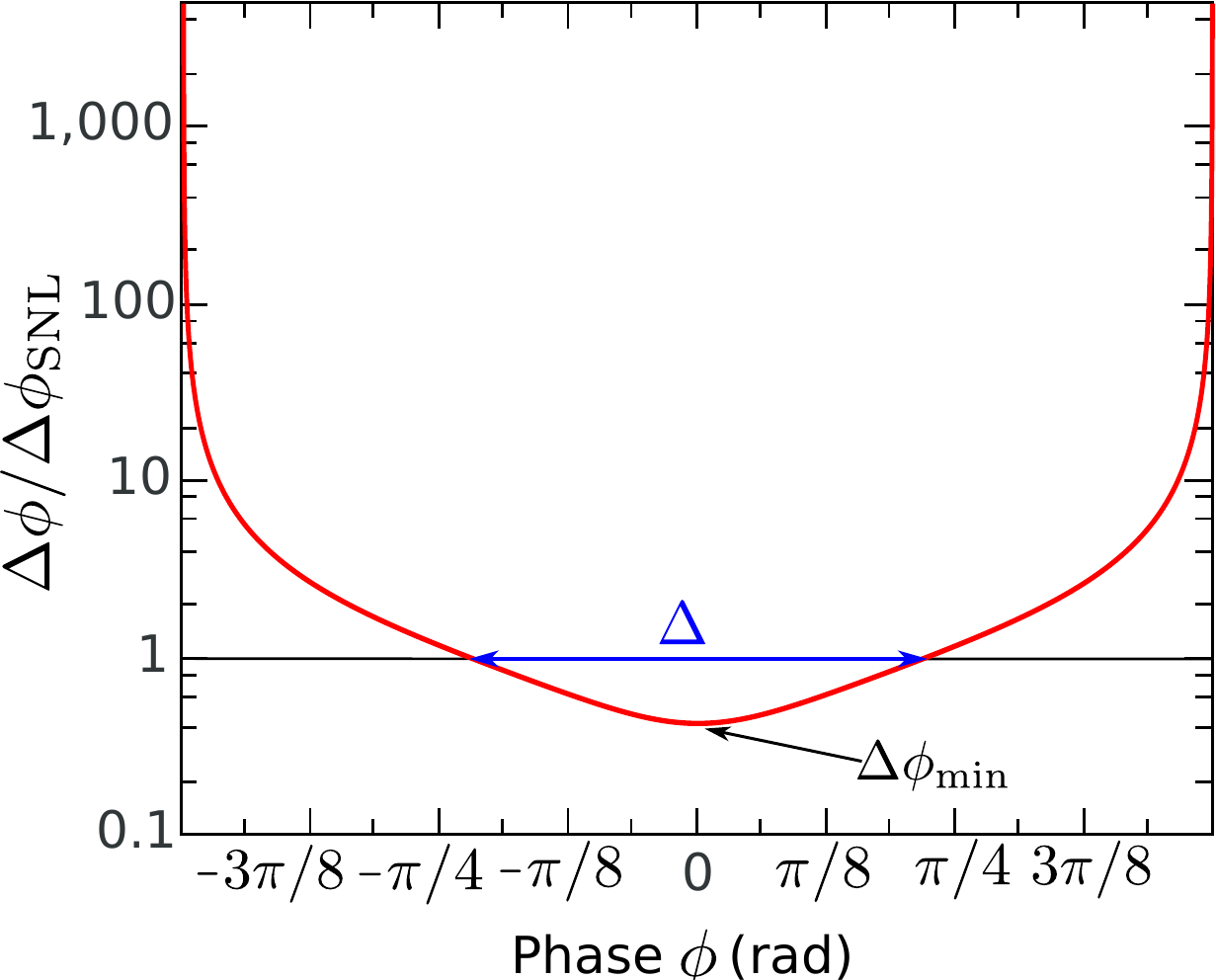}
  \caption{Phase sensitivity of an SU(2) interferometer preceded by a squeezer and followed by an antisqueezer
(Fig.\ref{fig:linear}),  as a function of the phase shift in the interferometer, for the parameters given by
(\ref{params1}, \ref{params2}) and a homodyne detection. The best phase sensitivity $\Delta\phi_{\rm min}$ is obtained at
$\phi_0=0$. The supersensitive phase range $\Delta$, for which $\Delta\phi<\Delta\phi_{\rm SNL}$, is shown in
blue.}\label{BitsDefHomodyneDetection}
\end{figure}

Fig.~\ref{BitsDefHomodyneDetection} shows the phase sensitivity $\Delta\phi$ given by Eq.~\ref{lin:dPhi}, normalized to
the shot-noise-limited phase sensitivity $\Delta\phi_{\rm SNL}$, as a function of the phase $\phi$, for the following
set of parameters:
\begin{equation}\label{params1}
  r_1=1.15\,,\quad\mu=0.90\,,
\end{equation}
\begin{equation}\label{params2}
  |r_2|=3\,,\quad\eta=0.3\,.
\end{equation}
The horizontal line in Fig.~\ref{BitsDefHomodyneDetection} marks the SNL phase sensitivity. One can see from this plot
that for the values (\ref{params1}, \ref{params2}), which could be considered as ``reasonably optimistic'' ones,
$\Delta$ is indeed very close to $\pi/2$.

\begin{figure}
\centering \includegraphics[scale=0.55]{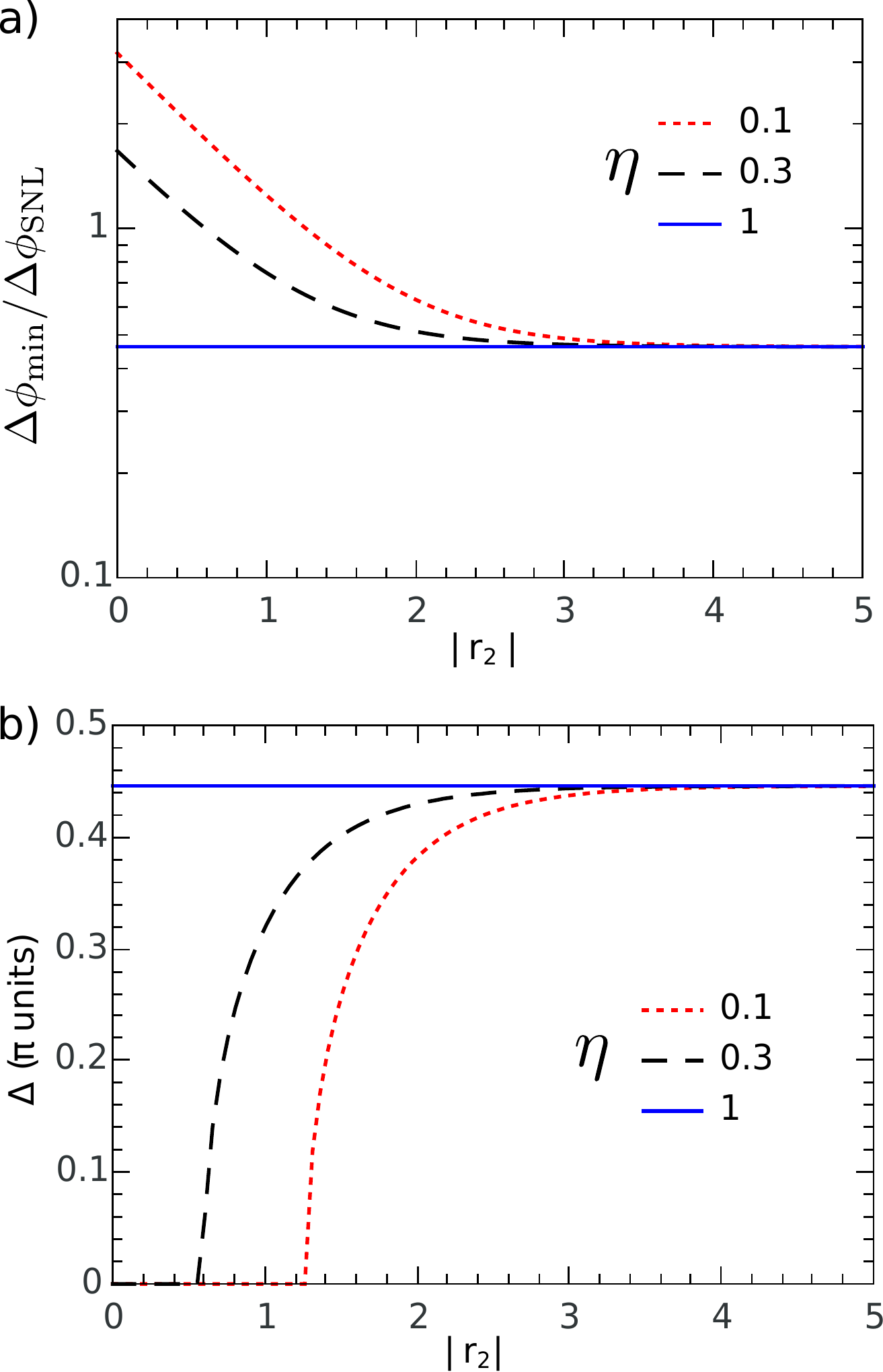}
\caption{ Optimal sensitivity $\Delta\phi_{\rm min}$
normalized to $\Delta\phi_{\rm SNL}$ (a) and the supersensitive phase range (b) of an SU(2) interferometer preceded by a
squeezer and followed by an antisqueezer as functions of the gain $r_2$ of the second amplifier for various values of
the detection efficiency $\eta$: blue line $\eta=1$,  black dashed line $\eta=0.3$ and red dotted line $\eta=0.1$. The
gain of the first amplifier $r_1$ and the internal transmission $\mu$ are given by \eqref{params1}.}
\label{LinearInterferometer}
\end{figure}

Now consider the dependence of the best phase sensitivity $\Delta\phi_{\rm min}$ \eqref{lin:dPhi_min} and the
supersensitive phase range $\Delta$ \eqref{Delta} on $r_2$ and $\eta$. In Fig.\ref{LinearInterferometer}, the ratio
$\Delta\phi_{\rm min}/\Delta\phi_{\rm SNL}$ and $\Delta$ are plotted as functions of $r_2$ for various values of $\eta$,
from the extremely lossy case corresponding to $\eta=0.1$ to no external losses. The values of $r_1$ and $\mu$ are
given by \eqref{params1}. In all cases, the external losses can be overcome by increasing the parametric gain of the
output amplifier. The sensitivity corresponding to the ideal detection ($\eta=1$) case can be recovered if

\begin{equation}
   e^{-2|r_2|} \ll \frac{\eta}{1-\eta}(\mu e^{-2r_1} + 1-\mu),
\end{equation}
more external losses requiring more parametric gain as one can see in Fig.\ref{LinearInterferometer}.

Also, an increase in the detection losses leads to a smaller super-sensitive phase range until the super-sensitivity
eventually disappears. By increasing the gain of the second squeezer, one improves not only the sensitivity as seen
previously but also the supersensitive phase range $\Delta$. A supersensitive phase range as broad as in the case of
lossless detection can always be retrieved by increasing the parametric gain of the second amplifier.

\section{Seeded SU(1,1) interferometer with homodyne detection}\label{sec:SU(1,1)Seeded}

\subsection{Degenerate interferometer}

We now consider an SU(1,1) interferometer made of two cascaded OPAs. We start with the simpler degenerate case shown in
Fig.\,\ref{fig:SU(1,1)}a. We suppose here that a coherent seed beam is injected into the first DOPA, and homodyne
detection is used at the output of the interferometer.

\begin{figure}
  \centering
  \includegraphics[scale=0.5]{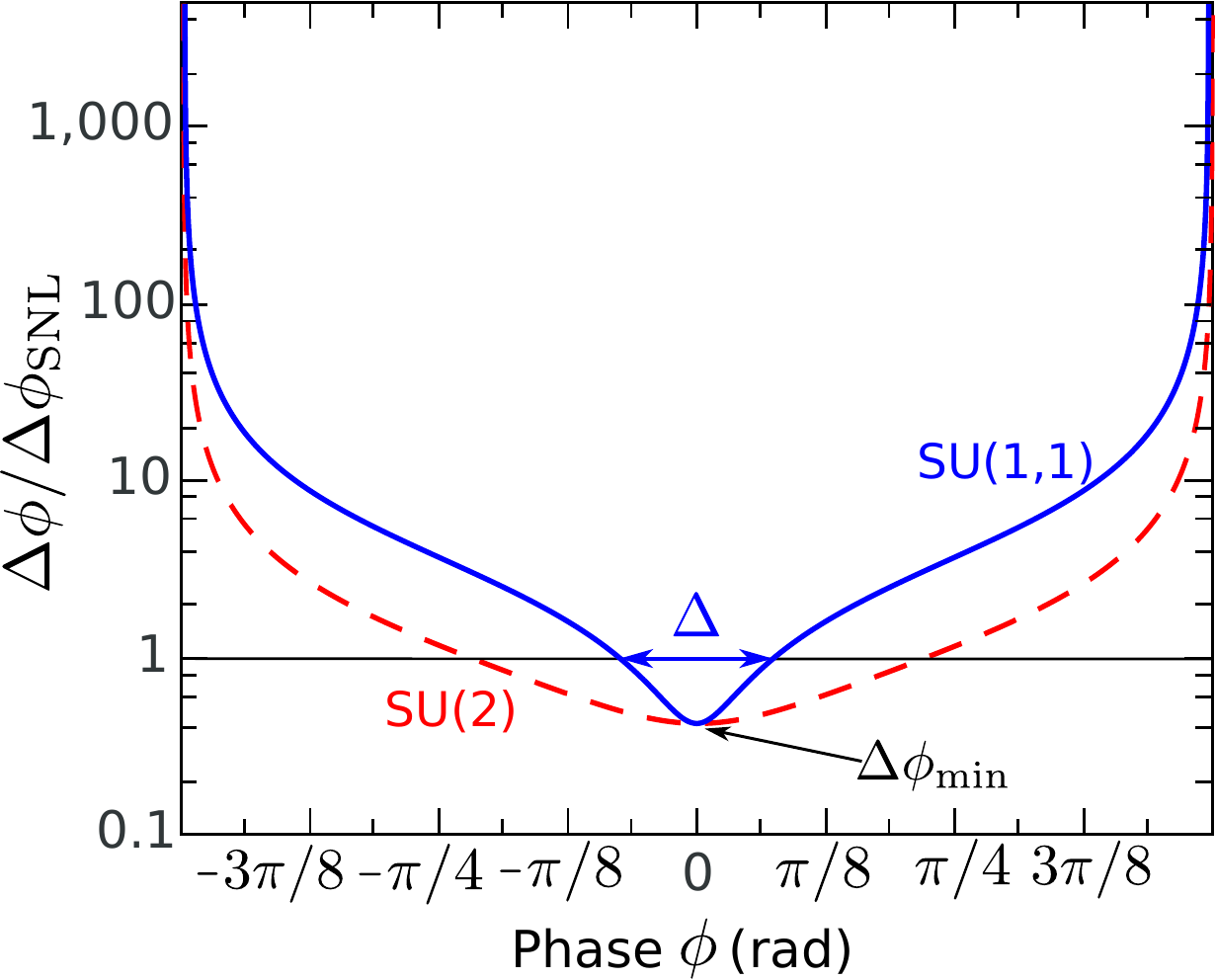}
  \caption{Phase sensitivity of a seeded SU(1,1) interferometer (blue line) with homodyne detection
(Fig.\ref{fig:SU(1,1)}), as a function of the phase shift in the interferometer, for the parameters from (\ref{params1},
\ref{params2}) and a homodyne detection. The best phase sensitivity $\Delta\phi_{\rm min}$ is obtained at $\phi_0=0$.
The
supersensitive phase
range $\Delta$ for which $\Delta\phi<\Delta\phi_{\rm SNL}$ is shown (blue arrow). For comparison, the phase
sensitivity of an SU(2) interferometer (red dashed line) from Fig.\ref{BitsDefHomodyneDetection} is plotted as
well.}\label{BitsDefHomodyneDetectionBis}
\end{figure}

Calculations given in \ref{nonlinear:homodyne} for a phase shift $\phi=\phi_s-\phi_p/2$ yield in this case that the
equation for the phase sensitivity again has the form \eqref{lin:dPhi}, with the same equation for  $\Delta\phi_{\rm
min}$ \eqref{lin:dPhi_min}, but with a different factor $K$:
\begin{equation}\label{NonLin:K}
  K = \frac{1}{4\alpha^2}
    \left(e^{2r_1} + \frac{1-\mu}{\mu} + \frac{1-\eta}{\mu\eta}\,e^{-2|r_2|}\right) .
\end{equation}
In all these equations for the SU(1,1) interferometer, $\alpha$ has the meaning of the classical amplitude inside the
interferometer, which is $e^{r_1}$ times stronger than the seed amplitude. The term $e^{2r_1}$ in \eqref{NonLin:K}
originates from the amplitude (anti-squeezed) light quadrature and noticeably reduces the supersensitive phase range of
this scheme in comparison with the linear interferometer, see Eq.\,\eqref{NonLin:Delta}.
Fig.\ref{BitsDefHomodyneDetectionBis} shows the phase sensitivity $\Delta\phi$ given by Eq.~\eqref{lin:dPhi}, with $K$
given by Eq.\eqref{NonLin:K}, as a function of the phase $\phi$. The sensitivity is normalized to the shot-noise-limited
phase sensitivity $\Delta\phi_{\rm SNL}$ and the parameters from \eqref{params1} and \eqref{params2} are used.

The mean number of photons used for the measurement in this case is also described by equation \eqref{lin:N}. Therefore,
the optimization of the best phase sensitivity $\Delta\phi_{\rm min}$ again gives the HL \eqref{min_dphi}.

\begin{figure}
\centering \includegraphics[scale=0.5]{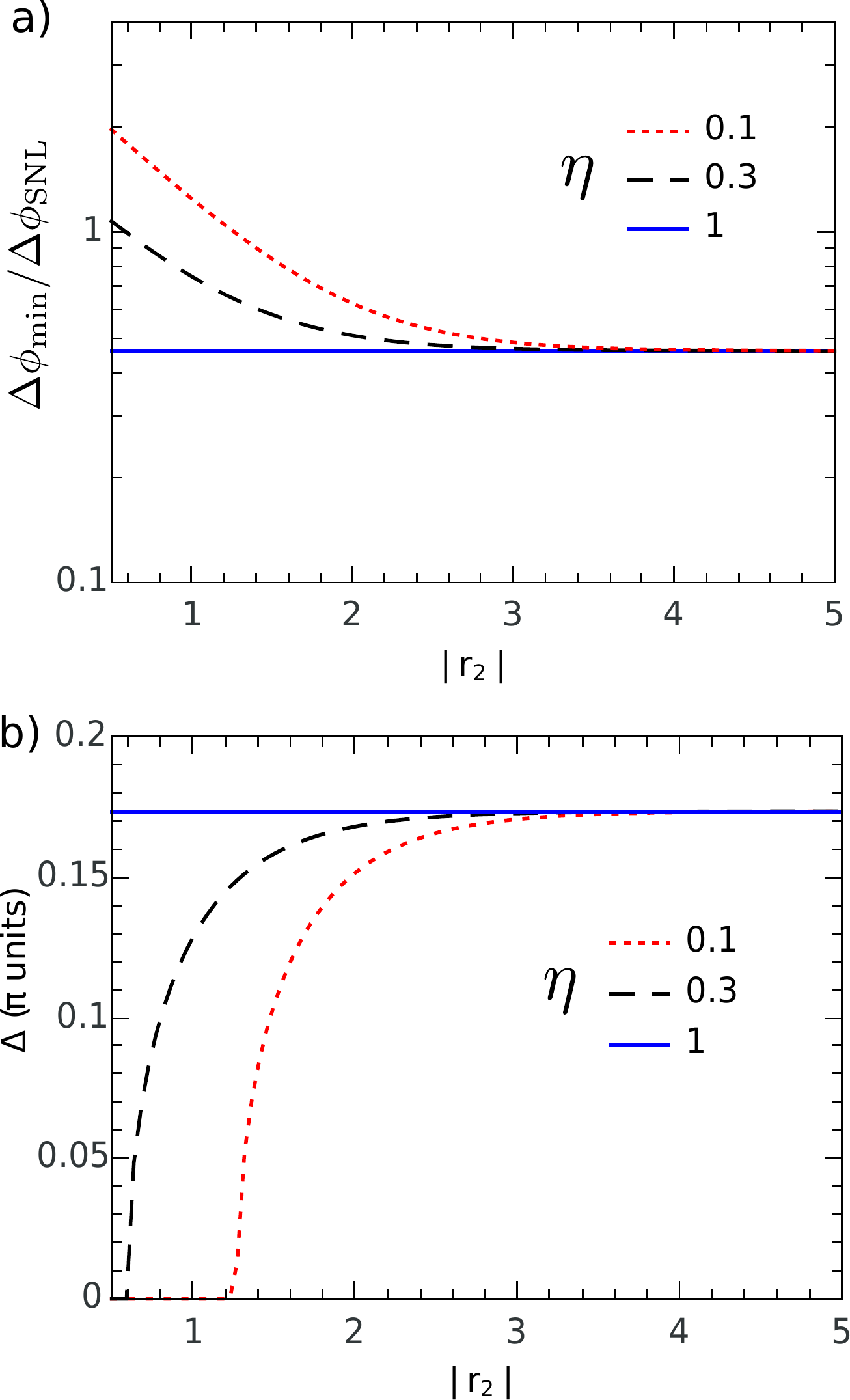}
\caption{Optimal phase sensitivity normalized to the SNL (a) and the supersensitive phase range (b) of a seeded
degenerate SU(1,1) interferometer with homodyne detection as functions of the gain $r_2$ of the second amplifier for
various values of the detection efficiency $\eta$: blue line $\eta=1$,  black dashed line $\eta=0.3$ and red dotted line
$\eta=0.1$. The gain of the first amplifier $r_1$ and the internal transmission
$\mu$ are given by \eqref{params1}.}
\label{NonLinSeeded}
\end{figure}

Using the same approach as in Sec.\,\ref{sec:Linear}, we can also calculate the asymptotic value of $\Delta$ for the
high-precision measurement case. It is easy to show that it is equal to

\begin{equation}\label{NonLin:Delta}
  \Delta \approx 2\arctan e^{-r_1} \approx 2e^{-r_1}\,
\end{equation}
i.e., it is $\sim e^{r_1}$ times narrower than in the linear interferometer case.

These results are summarized in Fig.\,~\ref{NonLinSeeded} showing the phase sensitivity normalized to the shot-noise
level and  the supersensitive phase range as functions of $r_2$ for different values of $\eta$, with the other
parameters given by \eqref{params1}. These results are very similar to the ones calculated for an SU(2) interferometer.
The only difference is a narrower supersensitive phase range.

\subsection{Non-degenerate interferometer}

Consider now a non-degenerate SU(1,1) interferometer as in Fig.\,\ref{fig:SU(1,1)}b  undergoing a
phase shift $\phi=(\phi_s + \phi_i -\phi_p)/2$. We show in \ref{app:nonlinear2} that if the signal and the idler
arms experience the same phase shifts and have the same optical losses, then this case is equivalent to the one of two
independent degenerate SU(1,1) interferometers, corresponding to the symmetric (+) and antisymmetric (-) optical modes
of the initial non-degenerate interferometer. Here, ``independent'' means that all optical fields in these modes are
uncorrelated. The squeeze factors of the two equivalent DOPAs in the symmetric mode are the same as the ones of the
initial NOPAs: $r_1$, $r_2$, respectively. For the antisymmetric-mode equivalent DOPAs, the squeeze factors are $-r_1$
and $-r_2$, respectively.

If two homodyne detectors are placed in the signal and idler output ports of the interferometer in
Fig.\,\ref{fig:SU(1,1)}b and they measure the same quadrature, then this equivalence can be extended to the detection
procedure as well. In this case, the sum and difference of the photocurrents of the signal and idler outputs
correspond to the output signals of the effective symmetric and antisymmetric non-degenerate interferometers. The
corresponding phase sensitivities, $\Delta\phi_+$ and $\Delta\phi_-$,
are given by equations (\ref{lin:dPhi},
\ref{lin:dPhi_min}, \ref{NonLin:K}) with inverted signs of the squeeze factors for the antisymmetric-mode
interferometer. Correspondingly, the total measurement error is
\begin{equation}
  (\Delta\phi)^2
  = \left[\frac{1}{(\Delta\phi_+)^2} + \frac{1}{(\Delta\phi_-)^2}\right]^{-1} .
\end{equation}

Let us find now the optimal distribution of the number of quanta between the signal and idler modes, assuming a given
total number of quanta. It follows from Eq.\,\eqref{n_b} that
\begin{equation}
  \alpha_s^2 + \alpha_i^2 = \alpha_+^2 + \alpha_-^2 \,,
\end{equation}
where $\alpha_{s,i}$ and
\begin{equation}
  \alpha_\pm = \frac{\alpha_s \pm \alpha_i}{\sqrt{2}}
\end{equation}
are the classical field amplitudes inside the interferometer in the respective
modes. It is evident that if $\Delta\phi_+<\Delta\phi_-$, then all power should be redistributed to the ``+'' mode,
giving $\alpha_-=0$, and if $\Delta\phi_+>\Delta\phi_-$, then to the ``--'' mode,  giving $\alpha_+=0$. In both cases,
the seed inputs have to be balanced: $|\alpha_s|^2=|\alpha_i|^2$, and in both cases, $\Delta\phi$ takes the form
of (\ref{lin:dPhi}, \ref{lin:dPhi_min}, \ref{NonLin:K}), with $\alpha^2$ corresponding to the total number of
quanta.

Experimentally, the ``+'' or ``--'' modes can also be accessed by mixing the two outputs of the interferometer on a
beamsplitter. The desired quadrature can then be measured by a single homodyne detector with the local oscillator
matching one of these modes at a given output of the beamsplitter.

\section{Seeded SU(1,1) interferometer with direct detection}\label{sec:SU(1,1)SeededDirect}

\begin{figure}[b]
  \centering
  \includegraphics[scale=0.5]{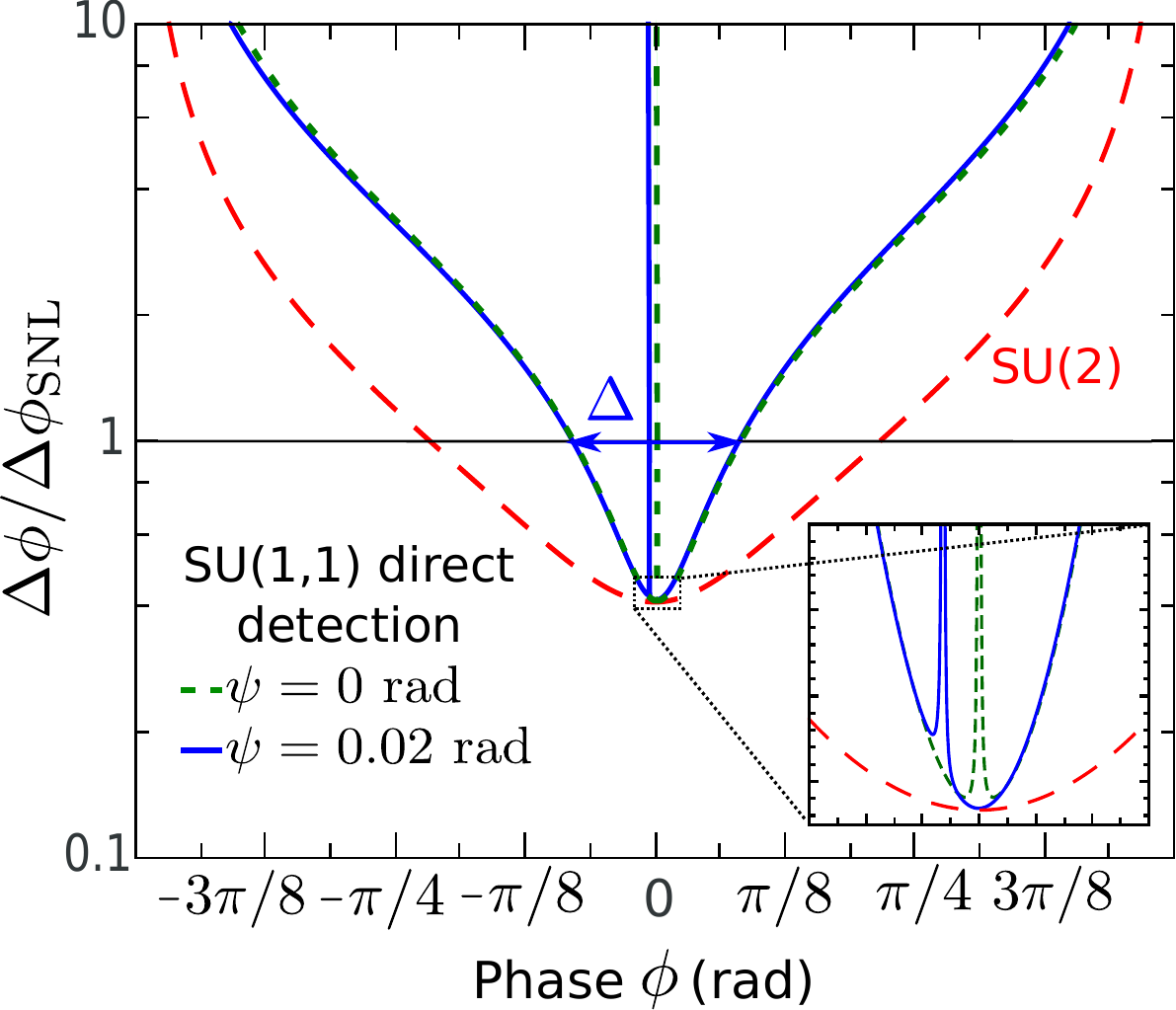}
  \caption{Phase sensitivity of a seeded SU(1,1) interferometer with direct detection (Fig.\ref{fig:SU(1,1)}), as a
function of the phase shift in the interferometer, for the parameters from (\ref{params1}, \ref{params2}) and a zoom
into the optimized sensitivity region around $\phi=0$ in the inset. Different seed phases $\psi$ are considered. For
comparison, the phase sensitivity of an SU(2) interferometer (red dashed line) with homodyne detection
from  Fig.\ref{BitsDefHomodyneDetection} is plotted as well. The best phase sensitivity
$\Delta\phi_{\rm min}$ is obtained at $\phi_0=0$ if the seed is slightly out of phase, $\psi\neq0$. The supersensitive
phase range $\Delta$, for which $\Delta\phi<\Delta\phi_{\rm SNL}$ is shown by a blue arrow.}\label{BitsDefDetectionTer}
\end{figure}

In many cases, the homodyne detection procedure considered above can be substituted by the simpler
measurement of the total output intensity, \textit{a direct detection} scheme. Unlike homodyne detection, it does not
require the use of a local oscillator, and it can be applied whenever the measured signal exceeds considerably the
detector dark noise. We show here that direct detection is able to provide a sensitivity close
to the one of the homodyne detection case. Here and in the next section we limit ourselves to the degenerate case only.

Similar to to the schemes considered above, we assume that the first and the second OPAs are tuned in anti-phase to each
other: $r_1>0, r_2<0$. However, we suppose that the seed phase inside the interferometer could differ from the squeeze
phase by some angle $\psi$, that is, the corresponding complex amplitude of the seed has the form of $\alpha
e^{-i\psi}$, where $\alpha$ is real.

The degenerate SU(1,1) with the direct detection is considered in \ref{nonlinear:seed_direct},
with an account for the above assumptions. The resulting equations (\ref{Dphi_direct_seeded}-\ref{dNdphi_direct_seeded})
for the phase sensitivity are quite cumbersome. Therefore, let us consider some characteristic
particular cases.

As in the previous sections, we start with the ideal lossless case $\mu=\eta=1$. Numerical optimization shows that
if $e^{-|r_{1,2}|}\ll1$, then the minimum of $\Delta\phi$ occurs at
\begin{equation}
  \phi = -(\phi+\psi) \approx \pm e^{-2|r_2|} \,,
\end{equation}
and is
\begin{equation}\label{OptPhaseSens}
  \Delta\phi_{\rm min} = \frac{e^{-r_1}}{2\alpha} \,.
\end{equation}
This coincides with the best sensitivity of the homodyne detection cases considered in the previous sections. The mean
number of photons used for the measurement in this case is again given by equation \eqref{lin:N}, therefore, the direct
detection scheme can also reach the HL \eqref{min_dphi}.

It follows from Eqs.\,(\ref{Dphi_direct_seeded}-\ref{dNdphi_direct_seeded}) that the best sensitivity is achieved at
small values of $\phi$. By assuming again $e^{-|r_{1,2}|}\ll1$, Eq.~\eqref{Dphi_direct_seeded} can be approximated as
follows:
\begin{equation}\label{Dphi_ds_app}
  \begin{array}{rcl}
    (\Delta\phi)^2 &=& \dfrac{1}{4\alpha^2}\Biggl\{
        e^{-2r_1} + \biggl(\phi + \dfrac{e^{-4|r_2|}}{\phi+\psi}\biggr)^2e^{2r_1}\\ 
        &  & + \dfrac{1-\mu}{\mu}\biggl[1 + \dfrac{e^{-8|r_2|}}{(\phi+\psi)^2}\biggr] \\
        &  & + \dfrac{1-\eta}{\mu\eta}e^{-2|r_2|}
            \biggl[1 + \dfrac{e^{-4|r_2|}}{(\phi+\psi)^2}\biggr]
      \Biggr\} .
  \end{array}
\end{equation}

Comparison of this equation with Eqs.\,(\ref{lin:dPhi}, \ref{lin:dPhi_min}, \ref{NonLin:K}) shows that direct
detection provides almost the same performance as homodyne detection and, in particular, results in almost
the same supersensitive phase range, approximately $e^{-r_1}$ smaller than in the SU(2) case:
\begin{equation}\label{NonLin:Delta-2}
  \Delta \approx2e^{-r_1} \,.
\end{equation}
Also, similar to the previous cases, if $|r_2|\to\infty$, then the term containing the external losses vanishes.

\begin{figure}[h!]
\centering \includegraphics[scale=0.55]{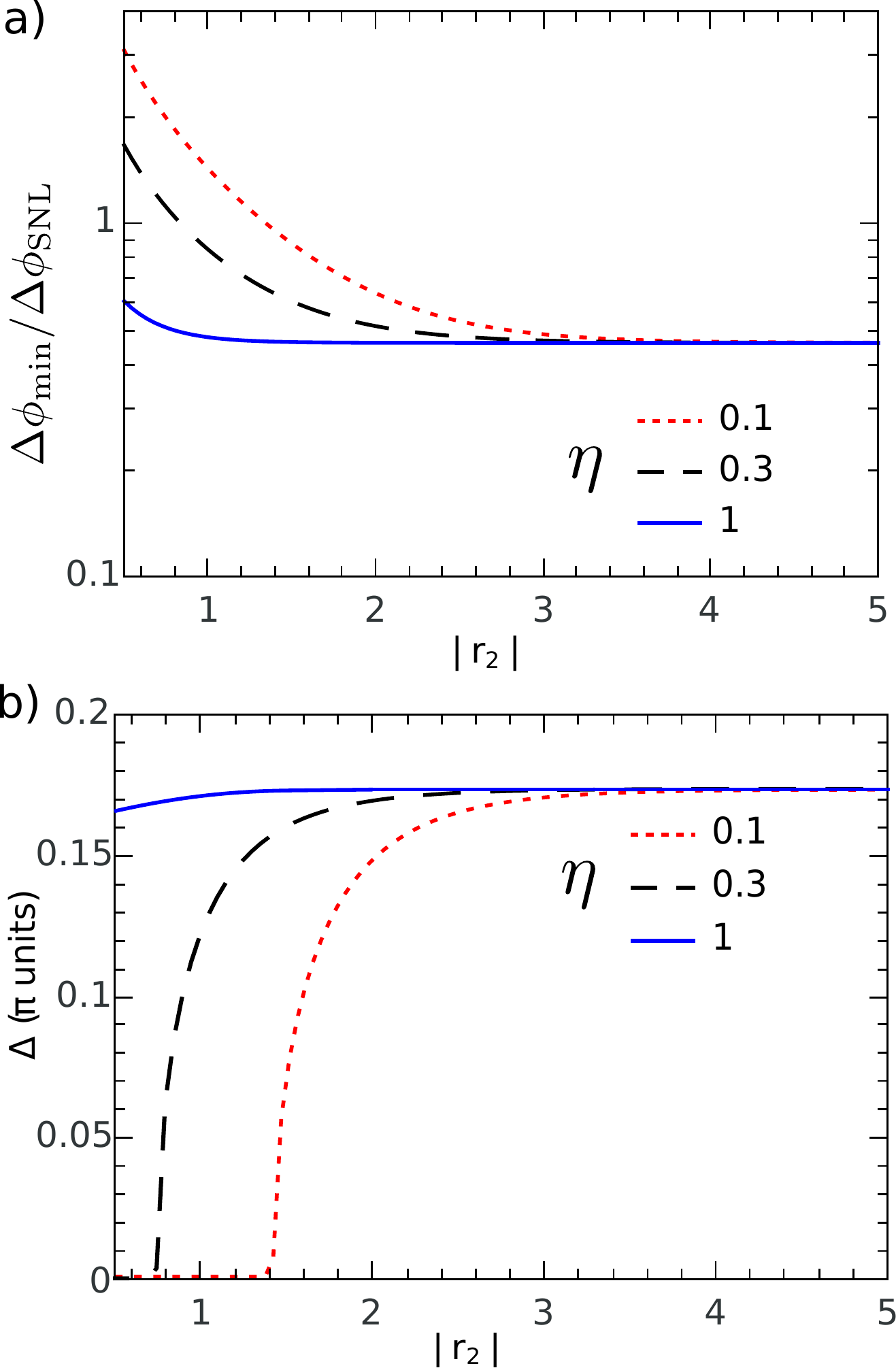}
\caption{Optimal phase sensitivity $\Delta\phi_{\rm min}$ normalized to $\Delta\phi_{\rm SNL}$ (a) and the
supersensitive phase range $\Delta$ (b) of a seeded degenerate SU(1,1) interferometer with direct detection as functions
of the gain $r_2$ of the second amplifier for various values of the detection efficiency $\eta$: blue line $\eta=1$,
black dashed line $\eta=0.3$ and red dotted line $\eta=0.1$. The gain of the first amplifier $r_1$ and the internal
transmission $\mu$ are given by \eqref{params1}.}
\label{NonLinSeeded-Int}
\end{figure}

However, function \eqref{Dphi_ds_app} has a narrow peak at $\phi+\psi\to0$, with the width $\sim e^{-2|r_2|}$,
originating from the absence of phase sensitivity at this point:
\begin{equation}
  \partd{\mean{N_f}}{\phi}\biggr|_{\phi+\psi=0} = 0 \,,
\end{equation}
see Eq.\,\eqref{dNdphi_direct_seeded}. This peak divides the supersensitive region into two
parts and can be considered as the main drawback of the direct detection scheme compared to the homodyne one.

These results are illustrated by Fig.\,~\ref{BitsDefDetectionTer} showing the phase sensitivity normalized to the
shot-noise level and by Fig.~\ref{NonLinSeeded-Int} showing the dependence of the optimal phase sensitivity
$\Delta\phi_{\rm min}$ and of the supersensitive phase range $\Delta$ on $r_2$ and $\eta$ for the parameters
(\ref{params1}).

In Fig.\,~\ref{BitsDefDetectionTer} we can see that the phase sensitivity in the direct detection scheme is much similar
to the SU(1,1) homodyne case shown in Fig.~\ref{BitsDefHomodyneDetectionBis} except for the extra peak denoting the
absence of sensitivity for $\phi+\psi\to0$ as explained earlier. In particular, the phase supersensitivity range is the
same as in the homodyne detection case when neglecting the small insensitive peak range.  The SU(2) homodyne case (red
dashed line) is also shown as a reference. Two curves are presented corresponding to  two values of the seed phase
$\psi$. By tuning the value of $\psi$ the insensitive peak can be moved away from $\phi=0$ as follows from the graph.
The optimal phase sensitivity $\Delta\phi_{\rm min}$  can be improved this way as visible in the inset of
Fig.\ref{BitsDefDetectionTer} which is a zoom into the $\phi=0$ region. For the particular case of the parameters
(\ref{params1}, \ref{params2}), the best phase sensitivity corresponds to $\psi\approx0.2$ and approaches the lossless
limit \eqref{OptPhaseSens}. In Fig.~\ref{NonLinSeeded-Int}a,b we observe the same trend as for the homodyne detection
case: by increasing the gain of the second amplifier one can compensate for the effect of external losses on the optimal
phase sensitivity and the supersensitivity phase range.

\section{Unseeded SU(1,1) interferometer with direct detection}\label{sec:SU(1,1)Unseeded}

Finally, we consider the simplest case of an unseeded SU(1,1) interferometer followed by a direct detection.
Calculations for this regime are made in \ref{nonlinear:direct}; the corresponding phase sensitivity is described by
Eqs.\,(\ref{nonlin_direct_dphi}-\ref{AB}), which have a rather sophisticated structure.

\begin{figure}[h!]
\centering \includegraphics[scale=0.5]{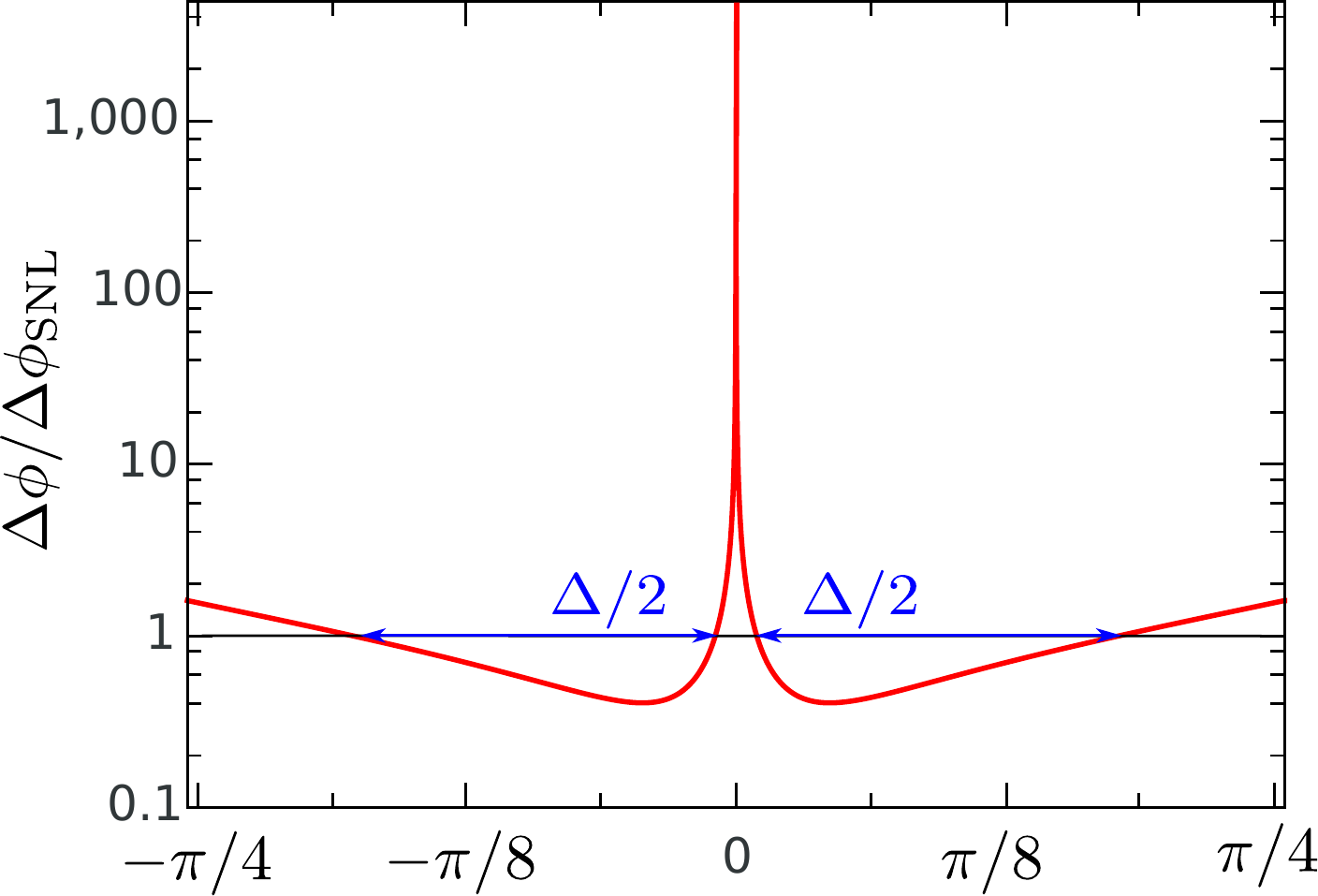}
\caption{Phase sensitivity of an unseeded degenerate SU(1,1) interferometer with direct detection as a function of the
phase shift for the parameters (\ref{params1}, \ref{params2}).The supersensitive phase range $\Delta$ is indicated in
the graph.}
\label{BitsDefDirectDetection}
\end{figure}

The dependence of the phase sensitivity \eqref{nonlin_direct_dphi} on the working point $\phi$ is shown in
Fig.~\ref{BitsDefDirectDetection}. The parameters of the interferometer, as in the previous cases, are given by
Eqs.\,(\ref{params1}, \ref{params2}). One can see that similar to the previous (seeded direct detection) case, the
supersensitive region is split into two parts by a narrow peak at $\phi=0$.

It can be also shown that for the strong squeezing case, the optimal working point $\phi_0$, which gives the best phase
sensitivity, is close to the dark fringe $\phi=0$, although does not exactly coincide with it. Therefore, suppose that
$|\phi|\ll1$. Assume also that $r_1>0$, $r_2<0$, which corresponds to the same ``squeezing/antisqueezing'' procedure
that we considered above. In this case, the phase sensitivity is
\begin{equation}\label{Delta_phi_direct}
  (\Delta\phi)^2 = \frac{1}{2}\biggl[
      \phi^2 + \frac{\cosh2R + A/2}{|\sinh2r_1\sinh2r_2|}
      + \frac{\sinh^22R + 2(A\sinh^2R + B)}{4\phi^2\sinh^22r_1\sinh^22r_2}
    \biggr] ,
\end{equation}
where
\begin{equation}
  R = r_1 + r_2
\end{equation}
and the phase-independent factors $A$, $B$ are defined by the the optical losses and the additional photon-number
measurement uncertainty $\Delta N_d$ introduced by the photocounter, see \eqref{AB}. This additional measurement uncertainty becomes
important only in the unseeded direct detection case, because in this case the number of photons measured by the
detector at the dark fringe is relatively small. For homodyne detection and/or seeded schemes, the contribution of this
noise can be made arbitrary small by simply increasing the power of the local oscillator and/or seed. Therefore, we did
not take it into account in the previous sections.

The minimum of $\Delta\phi$ in $\phi$ is achieved at
\begin{equation}\label{phi_min}
 \phi_0^2 = \frac{\sqrt{\sinh^22R + 2(A\sinh^2R + B)}}{2|\sinh2r_1\sinh2r_2|}
\end{equation}
and is equal to
\begin{equation}\label{Delta_phi_opt}
  (\Delta\phi_{\rm min})^2
  = \frac{\sqrt{\sinh^22R + 2(A\sinh^2R + B)} + \cosh2R + A/2}
      {2|\sinh2r_1\sinh2r_2|} \,.
\end{equation}

It is instructive to consider the asymptotic case of \eqref{Delta_phi_direct} for very strong second (anti)squeezing,
$e^{|r_2|}\to\infty$. It is easy to see that in this case
\begin{equation}
  A \to \frac{1}{2}\,\frac{1-\mu}{\mu}\,e^{2|r_2|} \,, \qquad
B \to \frac{1}{8}\left(\frac{1-\mu}{\mu}\right)^2e^{4|r_2|} \,,
\end{equation}
and
\begin{equation}\label{Delta_phi_direct_asy}
  \begin{array}{rcl}
    (\Delta\phi)^2 &=& \dfrac{1}{2}\biggl\{
        \phi^2 + \dfrac{1}{\sinh2r_1}\biggl(e^{-2r_1} + \dfrac{1-\mu}{2\mu}\biggr) \\
      &&+ \dfrac{1}{4\phi^2\sinh^22r_1}
            \biggl[
                e^{-4r_1} + \dfrac{1-\mu}{\mu}\,e^{-2r_1}
                + \left(\dfrac{1-\mu}{\mu}\right)^2
              \biggr]
      \biggr\} .
  \end{array}
\end{equation}
Note that all terms originating from the output losses and the detector imperfections are absent in this equation.

\begin{figure}[h!]
\centering \includegraphics[scale=0.55]{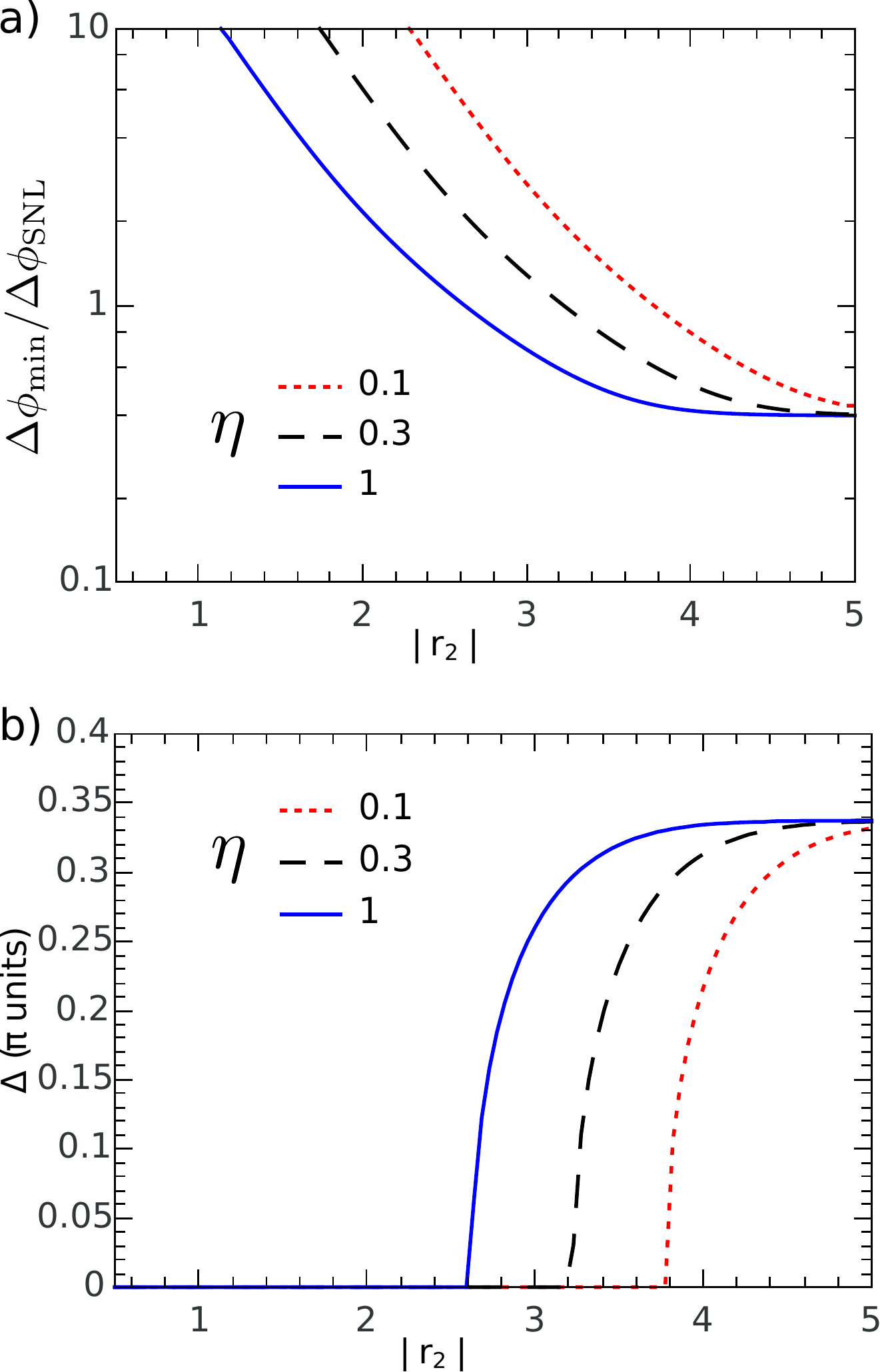}
\caption{Optimal phase sensitivity $\Delta\phi_{\rm min}$ normalized to $\Delta\phi_{\rm SNL}$ (a) and the
supersensitive phase range $\Delta$ (b) of an unseeded degenerate SU(1,1) interferometer with direct detection as
functions of the gain $r_2$ of the second amplifier for various values of the detection efficiency $\eta$: blue line
$\eta=1$,  black dashed line $\eta=0.3$ and red dotted line $\eta=0.1$. The gain of the first amplifier $r_1$ and the
internal transmission $\mu$ are given by \eqref{params1}; the detection noise is $\Delta N_d=100$~photons.}
\label{NonLinUnseeded-Noise}
\end{figure}

Fig.~\ref{NonLinUnseeded-Noise} presents the best phase sensitivity $\Delta\phi_{\rm min}$ of the interferometer and the
corresponding supersensitive  phase range as functions of the parametric gain of the second amplifier for the gain of
the first amplifier and the internal losses given by \eqref{params1}  and for $\Delta N_d=100$~photons. The optimal
phase sensitivity is not always obtained for the same working point, it depends on the gains of the amplifiers and
losses. It is nevertheless in the dark fringe region of the interferometer for which the photon noise is minimized (see
Fig.~\ref{BitsDefDirectDetection}). It is interesting to note that the balanced case $|r_1|=|r_2|=1.15$ does not provide
phase super-sensitivity even for $\eta=1$. However, the sensitivity improves as one increases the gain $r_2$ of the
second amplifier.

\section{Comparison between different interferometers and measurements schemes}\label{sec:Comparison}

\begin{figure}[h!]
\centering \includegraphics[scale=0.55]{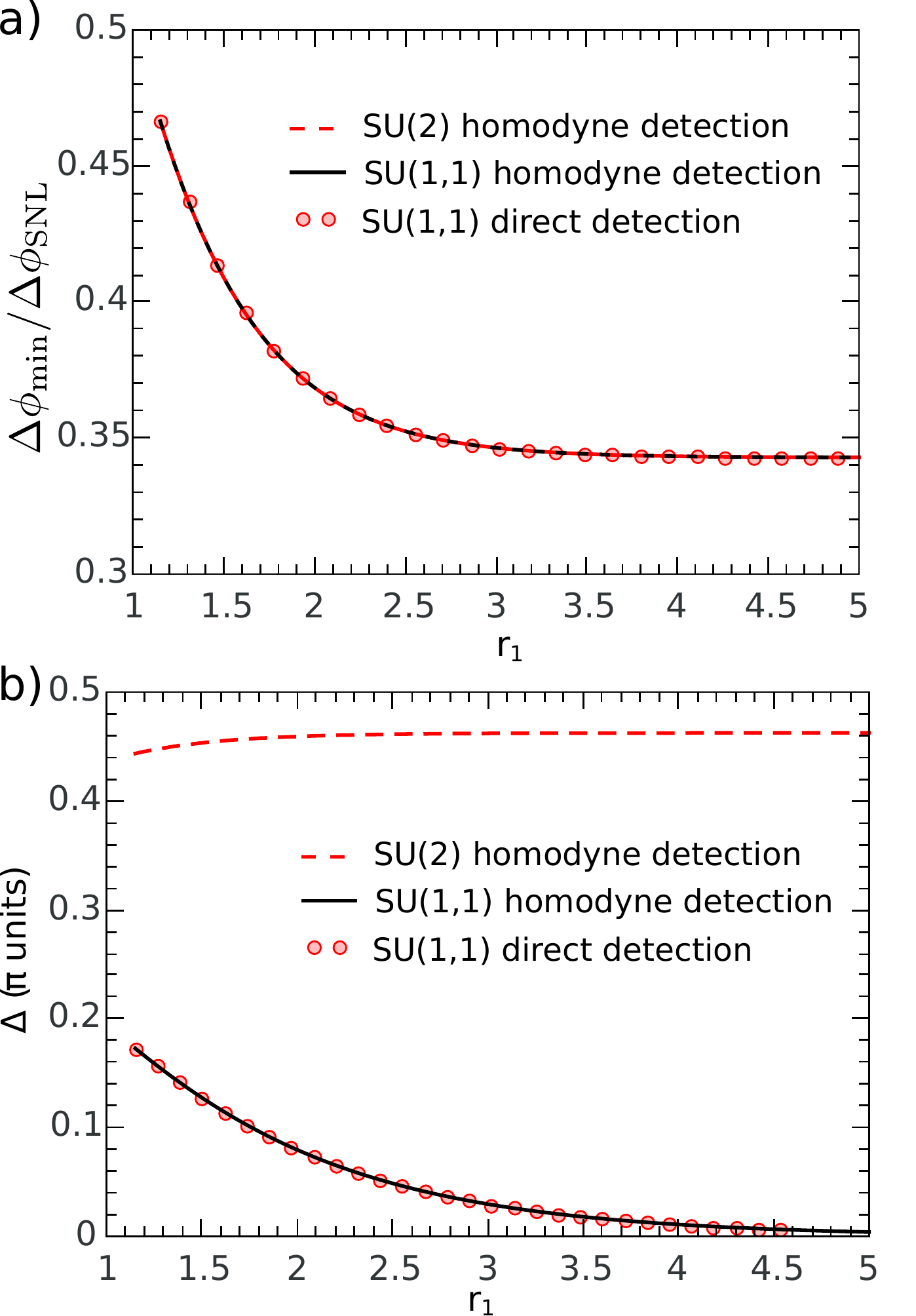}
\caption{Optimal phase sensitivity and the supersensitive phase range of a seeded SU(1,1) interferometer with direct
(red circles) and homodyne (solid  black line) detection compared to a seeded SU(2) interferometer (dashed red line).
The optimal sensitivity $\Delta\phi_{\rm min}$ normalized to $\Delta\phi_{\rm SNL}$ (a) and the supersensitive phase
range (b) versus the gain $r_1$ of the first amplifier.  The gain of the second amplifier is $|r_2|=3$, the internal
transmission $\mu=0.90$, and the external transmission $\eta=0.3$ in all cases. }
\label{InterferometerComparison3}
\end{figure}

Finally, in Fig.\ref{InterferometerComparison3} we summarize our findings and compare the different schemes studied
previously. In both panels, red dashed line represents the SU(2) interferometer with homodyne detection, black solid
line, the seeded SU(1,1) interferometer with homodyne detection, and red dotted line, the seeded SU(1,1) interferometer
with direct detection. In all cases, the gain of the second amplifier is $|r_2|=3$, the internal transmission $\mu=0.9$,
and the external transmission $\eta=0.3$.

Panel a shows  the best sensitivity $\Delta\phi_{\rm min}$ normalized to $\Delta\phi_{\rm SNL}$  and panel b, the
supersensitive phase range against the parametric gain of the first amplifier. As expected from
Eqs.~(\ref{lin:dPhi})-(\ref{lin:K}) and Eq.~(\ref{Dphi_ds_app}), the optimal
sensitivity $\Delta\phi_{\rm min}$ can be improved by increasing the gain of the first amplifier for both SU(1,1) and
SU(2) interferometers in all detection cases, see Fig.\ref{InterferometerComparison3}a.  Concerning the supersensitive
phase range in Fig.\ref{InterferometerComparison3}b, one can see that increasing the parametric gain of the first
amplifier degrades the supersensitive phase range for the SU(1,1) interferometer (red dots and solid
black line) conversely to the SU(2) interferometer. Indeed, the supersensitive phase range is dependent on $r_1$ in the
SU(1,1) case as already noticed in Section \ref{sec:SU(1,1)Seeded}, Eq.~(\ref{NonLin:Delta}).

\section{Conclusion}\label{sec:Conclusion}

We have studied the phase sensitivity properties of an SU(1,1) interferometer, considering the cases of both direct and
homodyne detection at the output and taking into account internal and external losses as well as the detector noise.
We have shown that the balanced configuration of an SU(1,1) interferometer, commonly
considered in the literature, in which the parametric gain values of both amplifiers are equal, is not the optimal one.
Increasing the gain of the second parametric amplifier always leads to a better sensitivity and a broader
super-sensitive phase range. At a given gain of the first amplifier, a sufficiently large gain of the second amplifier
can fully compensate for the detection losses. Although the gain unbalancing can be problematic for a FWM-based SU(1,1)
interferometer because of the mode mismatch, it can be realized with high-gain parametric down
conversion.

The `standard' configuration of an SU(1,1) interferometer, as proposed originally and used further in experiments, is
based on two non-degenerate parametric amplifiers. We have shown that its operation is similar to the one of two
independent degenerate SU(1,1) interferometers. In an experiment, the non-degenerate configuration is equivalent to the
degenerate one as long as one measures the sum of signal and idler parameters (quadratures or photon numbers).

We have also considered the unbalanced configuration of a linear (SU(2)) interferometer preceded by a squeezer and
followed by an anti-squeezer. This case can be particularly interesting for gravitational-wave detectors where an
existing SU(2) interferometer with squeezed input can be additionally equipped with an anti-squeezer at the output. We
have shown that an increase in the parametric gain of this additional anti-squeezer will considerably improve the phase
sensitivity. In particular, at any value of external (detection) losses, a sufficient gain of this anti-squeezer will
allow one to retrieve the phase sensitivity corresponding to the case of lossless detection.

\section{Acknowledgements}\label{sec:Acknowledgements}
This work was supported by the joint DFG-RFBR project CH $1591/2-1/16-52-12031$ NNIO a. The work of F.K. was supported by LIGO NSF Grant No
PHY-$1305863$ and Russian Foundation for Basic Research Grants No. $14-02-00399$ and $16-52-10069$.

\appendix

\section{Notations and the quadrature operators}

The annihilation operators are denoted by Roman letters $\hat{{\rm a}}, \hat{{\rm b}}\,,\ \etc$. The corresponding cosine and sine
quadrature operators \cite{Caves1985, Schumaker1985} are defined as
\begin{equation}
  \hat{{\rm a}}^c = \frac{\hat{{\rm a}} + \hat{{\rm a}}^\dagger}{\sqrt{2}} \,, \qquad
  \hat{{\rm a}}^s = \frac{\hat{{\rm a}} - \hat{{\rm a}}^\dagger}{i\sqrt{2}} \,.
\end{equation}
The two-component quadrature vectors are denoted by the boldface Roman letters:
\begin{equation}
  \hat{{\bf a}} = \svector{\hat{{\rm a}}^c}{\hat{{\rm a}}^s} ,
\end{equation}

We assume that all incident fields are in the coherent or vacuum state. In this case, their cosine
and
sine quadratures are uncorrelated noises with the uncertainties equal to 1/2.

The single-mode (degenerate) Bogolyubov (squeezing) transformation in the particular case of a real squeeze factor,
\begin{equation}
  \hat{{\rm b}} = \hat{{\rm a}}\cosh r + \hat{{\rm a}}^\dagger\sinh r,
\end{equation}
in terms of the quadrature operators has a simple form
\begin{equation}
  \hat{{\bf b}} = \mathbb{S}(r)\hat{{\bf a}} \,,
\end{equation}
\begin{equation}
  \mathbb{S}(r) = \smatrix{e^{r}}{0}{0}{e^{-r}} .
\end{equation}

\noindent The two-mode (non-degenerate) squeezing transformation has the form
\begin{equation}
  \hat{{\rm b}}_s = \hat{{\rm a}}_s\cosh r + \hat{{\rm a}}_i^\dagger\sinh r \,, \qquad
  \hat{{\rm b}}_i = \hat{{\rm a}}_i\cosh r + \hat{{\rm a}}_s^\dagger\sinh r \,,
\end{equation}
where s,i stand for the ``signal'' and ``idler'' modes. By introducing the symmetric and antisymmetric modes,
\begin{equation}
  \hat{{\rm a}}_+ = \frac{\hat{{\rm a}}_s + \hat{{\rm a}}_i}{2} \,, \qquad
  \hat{{\rm a}}_- = \frac{\hat{{\rm a}}_s - \hat{{\rm a}}_i}{2} \,,
\end{equation}
and similarly for $\hat{{\rm b}}$, it can be reduced to two independent single-mode transformations:
\begin{equation}
  \hat{{\rm b}}_\pm = \hat{{\rm a}}_\pm\cosh r \pm \hat{{\rm a}}_\pm^\dagger\sinh r \,.
\end{equation}

\noindent The phase shift transformation,
\begin{equation}
  \hat{{\rm b}} = \hat{{\rm a}}e^{-i\phi},
\end{equation}
correspondingly, has the form
\begin{equation}
  \hat{{\bf b}} = \mathbb{O}(\phi)\hat{{\bf a}} \,,
\end{equation}
where
\begin{equation}
  \mathbb{O}(\phi) = \smatrix{\cos\phi}{\sin\phi}{-\sin\phi}{\cos\phi}
  = \mathbb{I}\cos\phi - \mathbb{Y}\sin\phi \,,
\end{equation}
\begin{equation}
  \mathbb{I} = \smatrix{1}{0}{0}{1} , \qquad \mathbb{Y} = \smatrix{0}{-1}{1}{0}.
\end{equation}

\begin{figure*}
  \centering\includegraphics[width=0.7\textwidth]{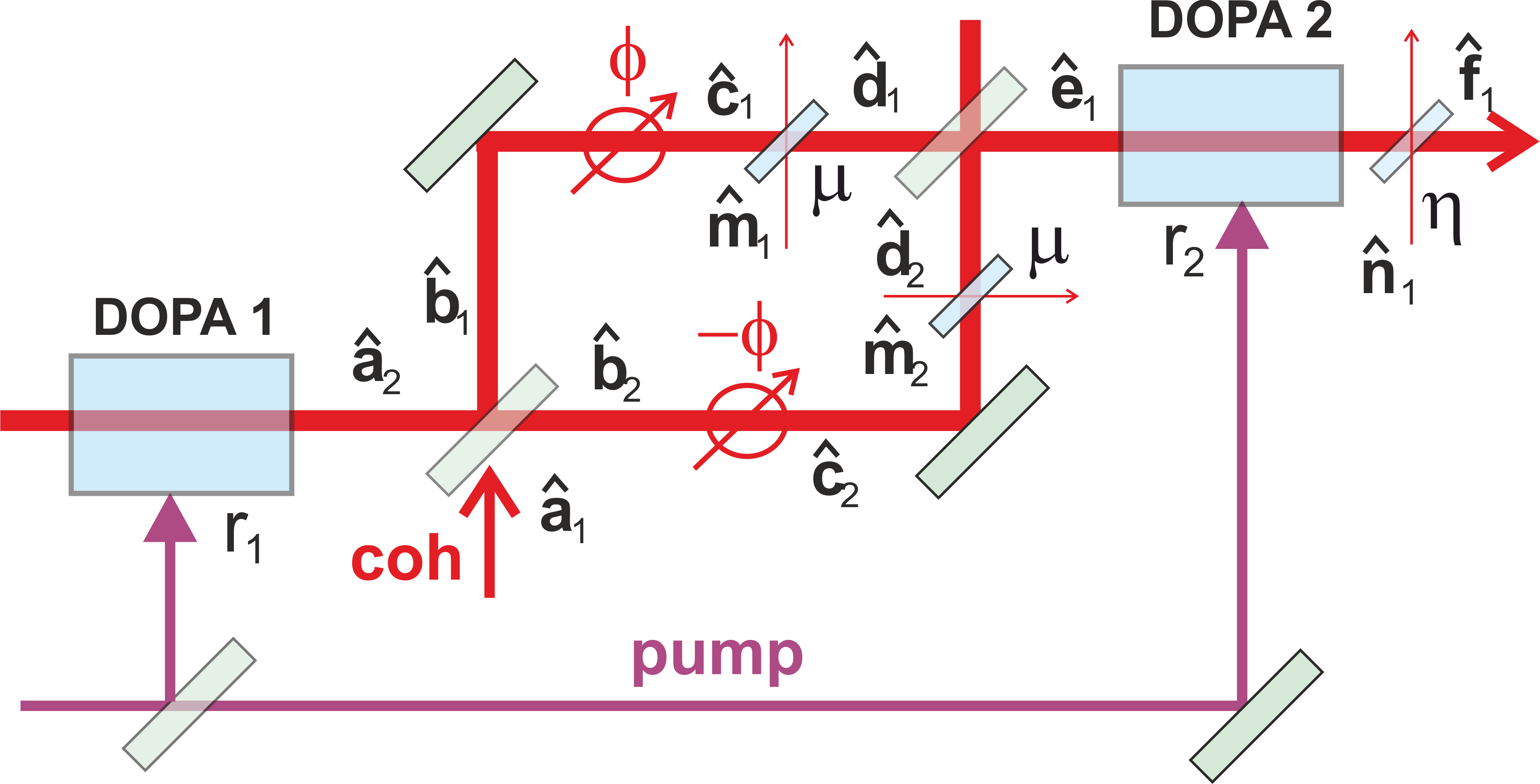}
  \caption{A linear interferometer preceded by a squeezer (degenerate optical parametric amplifier, DOPA1, with a
squeeze factor of $r_1$, and followed by an anti-squeezer (DOPA2), with the squeeze factor $r_2$. The figure shows the
notation used in the calculations. The pumping is depicted schematically.} \label{fig:linear_app}
\end{figure*}

\section{Linear interferometer}\label{app:linear}

\subsection{Field transformations in the interferometer}\label{lin:fields}

The two-component quadrature vectors of the incident fields (see Fig.~\,\ref{fig:linear_app}) are $\hat{{\bf a}}_1$, $\hat{{\bf a}}_2$.
The first beamsplitter transforms them into
\begin{equation}
  \hat{{\bf b}}_1 = \frac{\hat{{\bf a}}_1 + \hat{{\bf a}}_2}{\sqrt{2}}\,, \qquad
  \hat{{\bf b}}_2 = \frac{\hat{{\bf a}}_1 - \hat{{\bf a}}_2}{\sqrt{2}}\,.
\end{equation}

The phase shift gives, correspondingly,
\begin{equation}
  \hat{{\bf c}}_1 = \frac{\mathbb{O}(\phi)(\hat{{\bf a}}_1 + \hat{{\bf a}}_2)}{\sqrt{2}}\,,
    \qquad
  \hat{{\rm c}}_2
    = \frac{\mathbb{O}(-\phi)(\hat{{\bf a}}_1 - \hat{{\bf a}}_2)}{\sqrt{2}} \,.
\end{equation}

\noindent We model internal losses by an imaginary beamplitter with the power transmissivity $\mu$, which gives
\begin{equation}
  \begin{array}{l}
    \hat{{\bf d}}_1
      = \sqrt{\frac{\mu}{2}}\,\mathbb{O}(\phi)(\hat{{\bf a}}_1 + \hat{{\bf a}}_2)
        + \sqrt{1-\mu}\,\hat{{\bf m}}_1 \,, \\
    \hat{{\bf d}}_2
      =\sqrt{\frac{\mu}{2}}\,\mathbb{O}(-\phi)(\hat{{\bf a}}_1 - \hat{{\bf a}}_2)
        + \sqrt{1-\mu}\,\hat{{\bf m}}_2 \,,
  \end{array}
\end{equation}
\noindent where $\hat{{\bf m}}_{1,2}$ are the corresponding introduced vacuum noises.

\noindent At one of the outputs of the second beamsplitter, we have

\begin{equation}
  \hat{{\bf e}}_1 = \frac{\hat{{\bf d}}_1 - \hat{{\bf d}}_2}{\sqrt{2}}
  = \sqrt{\mu}(\hat{{\bf a}}_2\cos\phi - \mathbb{Y}\hat{{\bf a}}_1\sin\phi)
    + \sqrt{1-\mu}\,\hat{{\bf m}}_- ,
\end{equation}
where
\begin{equation}
  \hat{{\bf m}}_- = \frac{\hat{{\bf m}}_1 - \hat{{\bf m}}_2}{\sqrt{2}} \,.
\end{equation}
Finally, the second squeezer and the output losses give

\begin{equation}
  \begin{array}{rcl}
\hat{{\bf f}}_1
    &=&\sqrt{\eta}\,\mathbb{S}(r_2)\hat{{\bf e}}_1 + \sqrt{1-\eta}\,\hat{{\bf n}}_1 \\
    &=&\sqrt{\eta}\,\mathbb{S}(r_2)\bigl[
        \sqrt{\mu}(\hat{{\bf a}}_2\cos\phi - \mathbb{Y}\hat{{\bf a}}_1\sin\phi)
        + \sqrt{1-\mu}\,\hat{{\bf m}}_-
      \bigr] \\
    &&+ \sqrt{1-\eta}\,\hat{{\bf n}}_1 \,,
  \end{array}
\end{equation}

\noindent where $\eta$ is the power transmissivity of the imaginary beamsplitter that models the output losses, and
$\hat{{\bf n}}_1$ is the added vacuum noise.

\subsection{Homodyne detection}

Let there be coherent light at input 1, and squeezed vacuum at input 2:
\begin{equation}
  \begin{array}{rcl}
    \hat{{\bf a}}_1 &=& \svector{\sqrt{2}\,\alpha}{0} + \hat{{\bf z}}_1 \,, \\
    \hat{{\bf a}}_2 &=& \mathbb{S}(r_1)\hat{{\bf z}}_2 \,,
  \end{array}
\end{equation}
where $\hat{{\bf z}}_{1,2}$ are vacuum fields. In this case,
\begin{equation}\label{lin:f_2^c}
  \begin{array}{rcl}
    \hat{{\rm f}}_1^c &=& \sqrt{\mu\eta}\bigl[
          \hat{{\rm z}}_2^ce^{r_1}\cos\phi + \hat{{\rm z}}_1^s\sin\phi
        \bigr]e^{r_2} \\
      &&+ \sqrt{(1-\mu)\eta}\,\hat{{\rm m}}_-^ce^{r_2}
        + \sqrt{1-\eta}\,\hat{{\rm n}}_1^c,
  \end{array}
\end{equation}
\begin{equation}\label{lin:f_2^s}
  \begin{array}{rcl}
    \hat{{\rm f}}_1^s &=& \sqrt{\mu\eta}\bigl[
          \hat{{\rm z}}_2^se^{-r_1}\cos\phi
          - (\sqrt{2}\,\alpha + \hat{{\rm z}}_1^c)\sin\phi
        \bigr]e^{-r_2} \\
      &&+ \sqrt{(1-\mu)\eta}\,\hat{{\rm m}}_-^se^{-r_2}
        + \sqrt{1-\eta}\,\hat{{\rm n}}_1^s.
  \end{array}
\end{equation}
One can see that if $|\phi|\ll1$, then the sine quadrature contains the most significant part of the phase  information
(the term $\sqrt{2}\,\alpha\sin\phi$). Therefore, assume that it is this quadrature that is measured by a homodyne
detector. It follows from \eqref{lin:f_2^s} that the mean value and the uncertainty of $\hat{{\rm f}}_1^s$ are

\begin{equation}
  \mean{\hat{{\rm f}}_1^s} = -\sqrt{2\mu\eta}\,\alpha e^{-r_2}\sin\phi \,,
\end{equation}
\begin{equation}
  \begin{array}{rcl}
    (\Delta\hat{{\rm f}}_1^s)^2
    &=& \mean{(\hat{{\rm f}}_1^s - \mean{\hat{{\rm f}}_1^s})^2} \\
    &=& \frac{1}{2}\bigl[
            \mu\eta(e^{-2r_1}\cos^2\phi + \sin^2\phi)e^{-2r_2}
           + (1-\mu)\eta e^{-2r_2} \\
         &&+ 1-\eta
          \bigr] .
  \end{array}
\end{equation}

\noindent The phase measurement error is defined by

\begin{equation}
  (\Delta\phi)^2 = \frac{(\Delta\hat{{\rm f}}_1^s)^2}
      {\left(\partd{\mean{\hat{{\rm f}}_1^s}}{\phi}\right)^2} \,,
\end{equation}
which gives Eqs.\,(\ref{lin:dPhi}-\ref{lin:K}).

\begin{figure*}
\centering \includegraphics[scale=0.4]{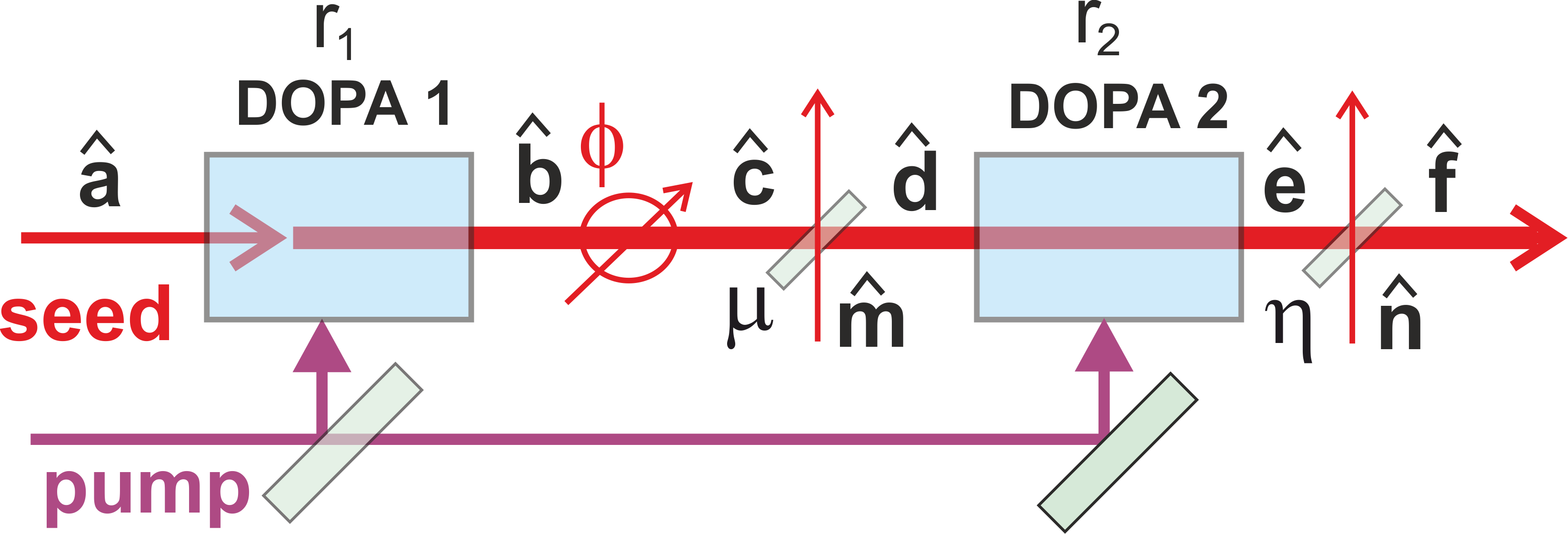}
\caption{Degenerate SU(1,1) interferometer and the notation used in the calculations.  The pumping of the DOPAs is shown schematically.}
\label{fig:DOPA}
\end{figure*}

\section{Degenerate SU(1,1) interferometer}\label{app:nonlinear}

\subsection{Field transformations}\label{nonlin:fields}

Here, we repeat the calculations of \ref{lin:fields} for the case of a degenerate SU(1,1)
interferometer. The scheme with the main notation is shown in Fig.~\ref{fig:DOPA}.

The two-components quadrature vector $\hat{{\bf a}}$ for the incident field becomes, after the first DOPA,
\begin{equation}\label{bf_b}
  \hat{{\bf b}} = \mathbb{S}(r_1)\hat{{\bf a}} \,.
\end{equation}
After the signal phase shift, it becomes
\begin{equation}
  \hat{{\bf c}} = \mathbb{O}(\phi)\hat{{\bf b}} \,.
\end{equation}
The internal losses are taken into account by the effective beamsplitter transformation,
\begin{equation}
  \hat{{\bf d}} = \sqrt{\mu}\,\hat{{\bf b}}  + \sqrt{1-\mu}\,\hat{{\bf m}} \,.
\end{equation}
After the second DOPA, the quadrature vector becomes
\begin{equation}\label{deg_out}
  \hat{{\bf e}} = \mathbb{S}(r_2)\hat{{\bf d}} \,.
\end{equation}
The external losses are taken into account by the effective beamsplitter transformation,
\begin{equation}\label{bf_f}
  \begin{array}{rcl}
    \hat{{\bf f}} &=& \sqrt{\eta}\,\hat{{\bf e}} + \sqrt{1-\eta}\,\hat{{\bf n}} \\
    &=& \sqrt{\eta}\,\mathbb{S}(r_2)\bigl[
          \sqrt{\mu}\,\mathbb{O}(\phi)\mathbb{S}(r_1)\hat{{\bf a}}
          + \sqrt{1-\mu}\,\hat{{\bf m}}
        \bigr]
      + \sqrt{1-\eta}\,\hat{{\bf n}} \,.
  \end{array}
\end{equation}

In the case of a coherent seed beam at the input of the first DOPA (Fig.~\ref{fig:DOPA}), the quadrature vector has the form
\begin{equation}
  \hat{{\bf a}} = \bm\zeta + \hat{{\bf z}} \,,
\end{equation}
where $\hat{{\bf z}}$ is a vacuum field and $\bm\zeta$ is the seed quadratures vector. In this case,
\begin{equation}\label{bf_b_seed}
  \hat{{\bf b}} = \mathbb{S}(r_1)\hat{{\bf a}} \,.
\end{equation}
\begin{equation}\label{bf_f_seed}
  \hat{{\bf f}} = \sqrt{\eta}\,\mathbb{S}(r_2)\bigl\{
        \sqrt{\mu}\,\mathbb{O}(\phi)[\bm\alpha + \mathbb{S}(r_1)\hat{{\bf a}}]
        + \sqrt{1-\mu}\,\hat{{\bf m}}
      \bigr\}
    + \sqrt{1-\eta}\,\hat{{\bf n}} \,,
\end{equation}
where
\begin{equation}
  \bm\alpha = \mathbb{S}(r_1)\bm\zeta = {}|\alpha|\svector{\cos\psi}{-\sin\psi}
\end{equation}
is the seed quadratures vector inside the interferometer. (Please note that here, $\alpha$ depends on $r_1$.)

\subsection{Seeded case with homodyne detection}\label{nonlinear:homodyne}

Here we suppose that the phase of the coherent seed is equal to zero:
\begin{equation}
  \bm\alpha = \svector{\sqrt{2}\,\alpha}{0} \,.
\end{equation}
In this case,

\begin{equation}\label{nonlin:f^c}
  \begin{array}{rcl}
    \hat{{\rm f}}^c &=& \sqrt{\mu\eta}\bigl[
          (\sqrt{2}\,\alpha + \hat{{\rm z}}^ce^{r_1})\cos\phi
          + \hat{{\rm z}}^se^{-r_1}\sin\phi
        \bigr]e^{r_2} \\
      &&+ \sqrt{(1-\mu)\eta}\,\hat{{\rm m}}^ce^{r_2}
        + \sqrt{1-\eta}\,\hat{{\rm n}}^c \,,
  \end{array}
\end{equation}
\begin{equation}\label{nonlin:f^s}
  \begin{array}{rcl}
    \hat{{\rm f}}^s &=& \sqrt{\mu\eta}\bigl[
          \hat{{\rm z}}^se^{-r_1}\cos\phi
          - (\sqrt{2}\,\alpha + \hat{{\rm z}}^ce^{r_1})\sin\phi
        \bigr]e^{-r_2} \\
      &&+ \sqrt{(1-\mu)\eta}\,\hat{{\rm m}}^se^{-r_2}
        + \sqrt{1-\eta}\,\hat{{\rm n}}^s \,,
  \end{array}
\end{equation}
As in \ref{app:linear}, at $|\phi|\ll1$ the sine quadrature contains the
most significant part of the phase  information (the term $\sqrt{2}\,\alpha\sin\phi$). It follows from
\eqref{nonlin:f^s}
that the mean value and the variance of $\hat{{\rm f}}^s$ are
\begin{equation}
  \mean{\hat{{\rm f}}^s} = -\sqrt{2\mu\eta}\,\alpha e^{-r_2}\sin\phi \,,
\end{equation}
\begin{equation}
  \begin{array}{rcl}
    (\Delta\hat{{\rm f}}^s)^2
    &=& \mean{(\hat{{\rm f}}^s - \mean{\hat{{\rm f}}^s})^2} \\
    &=& \frac{1}{2}\bigl[
            \mu\eta(e^{-2r_1}\cos^2\phi + e^{2r_1}\sin^2\phi)e^{-2r_2}
           + (1-\mu)\eta e^{-2r_2} \\
         &&+ 1-\eta
          \bigr] .
  \end{array}
\end{equation}
The phase measurement error is defined by
\begin{equation}
  (\Delta\phi)^2 = \frac{(\Delta\hat{{\rm f}}^s)^2}
      {\left(\partd{\mean{\hat{{\rm f}}^s}}{\phi}\right)^2} \,,
\end{equation}
which gives Eqs.\,(\ref{lin:dPhi}, \ref{lin:dPhi_min}, \ref{NonLin:K}).

\subsection{Seeded case with direct detection}\label{nonlinear:seed_direct}

Let us rewrite Eq\,\eqref{bf_f_seed} back in the annihilation/creation operator notation:
\begin{equation}\label{f_direct_seed}
  \hat{{\rm f}} = \sqrt{\mu\eta}\theta(\phi) + \hat{{\rm f}}_{\rm fl} \,,
\end{equation}
where
\begin{equation}
  \theta(\phi) = \alpha e^{-i\phi}\cosh r_2 + \alpha^* e^{i\phi}\sinh r_2 \,,
\end{equation}
\begin{equation}
  \alpha = |\alpha|e^{-i\psi} \,,
\end{equation}
\begin{equation}\label{f_fl}
  \begin{array}{rcl}
    \hat{{\rm f}}_{\rm fl} &=& \sqrt{\mu\eta}
      \bigl[C(\phi)\hat{{\rm a}} + S(\phi)\hat{{\rm a}}^\dagger\bigr]
    + \sqrt{\eta(1-\mu)}(\hat{{\rm m}}\cosh r_2 + \hat{{\rm m}}^\dagger\sinh r_2) \\
    &&+ \sqrt{1-\eta}\,\hat{{\rm n}} \,,
  \end{array}
\end{equation}
\begin{equation}\label{C,S}
  \begin{array}{rcl}
    C(\phi) &=& \cosh r_1\cosh r_2\,e^{-i\phi} + \sinh r_1\sinh r_2\,e^{i\phi} \,,\\
    S(\phi) &=& \sinh r_1\cosh r_2\,e^{-i\phi} + \cosh r_1\sinh r_2\,e^{i\phi} \,.
  \end{array}
\end{equation}

The number of quanta on the detector, up to small second-order terms, is equal to
\begin{equation}
  \hat{N}_f = \hat{{\rm f}}^\dagger\hat{{\rm f}} = \mean{\hat{N}_f} + \delta\hat{N}_f \,,
\end{equation}
where
\begin{equation}
  \mean{\hat{N}_f} = \mu\eta|\theta(\phi)|^2
  = \mu\eta|\alpha|^2[\cosh2r_2 + \sinh2r_2\cos2(\phi+\psi)]\,,
\end{equation}
\begin{equation}
  \delta\hat{N}_f = \sqrt{\mu\eta}
    [\theta(\phi)\hat{{\rm f}}_{\rm fl}^\dagger + \theta^*(\phi)\hat{{\rm f}}_{\rm fl}]\,.
\end{equation}
are the mean number and the variance of $\hat{N}_f$. Therefore,
\begin{equation}
  (\Delta N_f)^2 = \mean{(\delta\hat{N}_f)^2} = \mu\eta|\alpha|^2
    \bigl[\mu\eta\sigma_a^2 + (1-\mu)\eta\sigma_m^2 + (1-\eta)\sigma_n^2\bigr] ,
\end{equation}
and
\begin{equation}\label{Dphi_direct_seeded}
  (\Delta\phi)^2 = \frac{(\Delta N_f)^2}{\left(\partd{\mean{N_f}}{\phi}\right)^2}
  = \frac
      {\sigma_a^2 + \dfrac{1-\mu}{\mu}\,\sigma_m^2 + \dfrac{1-\eta}{\mu\eta}\,\sigma_n^2}
      {4|\alpha|^2\sinh^22r_2\sin^22(\phi+\psi)} \,,
\end{equation}
where
\begin{equation}
  \sigma_n^2 = e^{2r_2}\cos^2(\phi+\psi) + e^{-2r_2}\sin^2(\phi+\psi) \,,
\end{equation}
\begin{equation}
  \sigma_m^2 = e^{4r_2}\cos^2(\phi+\psi) + e^{-4r_2}\sin^2(\phi+\psi) \,,
\end{equation}
\begin{equation}
  \begin{array}{rcl}
    \sigma_a^2 &=& \cosh2r_1\cosh4r_2 \\
      &&+ \sinh2r_1[\cosh4r_2\cos2\phi\cos2(\phi+\psi) + \sin2\phi\sin2(\phi+\psi)] \\
      &&+ [\cosh2r_1\cos2(\phi+\psi) + \sinh2r_1\cos2\phi]\sinh4r_2 \,,
  \end{array}
\end{equation}
and
\begin{equation}
  \partd{\mean{N_f}}{\phi} = -2\mu\eta|\alpha|^2\sinh2r_2\sin2(\phi+\psi) \,.
\end{equation}

Numerical optimization shows that if $e^{r_1}\gg1$, $e^{-r_2}\gg1$, then the optimal values of $\phi$, $\psi$ are close to 0. In this case,
\begin{equation}
  \sigma_n^2 = e^{-2|r_2|} + e^{2|r_2|}(\phi+\psi)^2 \,,
\end{equation}
\begin{equation}
  \sigma_m^2 = e^{-4|r_2|} + e^{4|r_2|}(\phi+\psi)^2 \,,
\end{equation}
\begin{equation}
  \sigma_a^2 = e^{-4|r_2|+2r_1} + 2e^{2r_1}\phi(\phi+\psi)
    + e^{4|r_2|}(e^{2r_1}\phi^2 + e^{-2r_1})(\phi+\psi)^2 \,, \label{sigma2a_app}
\end{equation}
and
\begin{equation}\label{dNdphi_direct_seeded}
  \partd{\mean{N_f}}{\phi} = 2\mu\eta|\alpha|^2e^{2|r_2|}(\phi+\psi) \,,
\end{equation}
The minimum of \eqref{sigma2a_app} in $\phi$ is at
\begin{equation}
  \phi = -\frac{e^{-4|r_2|}}{\phi+\psi}
\end{equation}
and is equal to
\begin{equation}
  \sigma_a^2 = e^{4|r_2|-2r_1}(\phi+\psi)^2 \,.
\end{equation}

\subsection{Unseeded case with direct detection}\label{nonlinear:direct}

If $\alpha=0$, then $\hat{{\rm f}}=\hat{{\rm f}}_{\rm fl}$ [see Eq.\,\eqref{f_fl}], and the mean value and the variance
of the number of quanta at the output are equal to

\begin{equation}\label{N}
  \mean{\hat{N}_f} = \mean{\hat{{\rm f}}^\dagger\hat{{\rm f}}}
  = \eta\left[\mu|S(\phi)|^2 + (1-\mu)\sinh^2r_2\right] ,
\end{equation}
\begin{equation}
  \begin{array}{rcl}
    (\Delta N_f)^2 &=& \mean{\hat{N}_f^2} - \mean{\hat{N}_f}^2 \\
    &=& \eta^2\bigl\{
            \mu^2|S(\phi)|^2\bigl[|C(\phi)|^2 + |S(\phi)|^2\bigr] \\
      &&    + 2\mu(1-\mu)|S(\phi)|^2\sinh^2r_2
            + (1-\mu)^2\sinh^2r_2\cosh2r_2
         \bigr\} \\
      &&  + \mean{\hat{N}_f} \,.
  \end{array}
\end{equation}

Due to the losses in a realistic interferometer, as well as this additional measurement error, a good phase sensitivity
can be evidently achieved only if

\begin{figure*}
  \centering\includegraphics[width=0.7\textwidth]{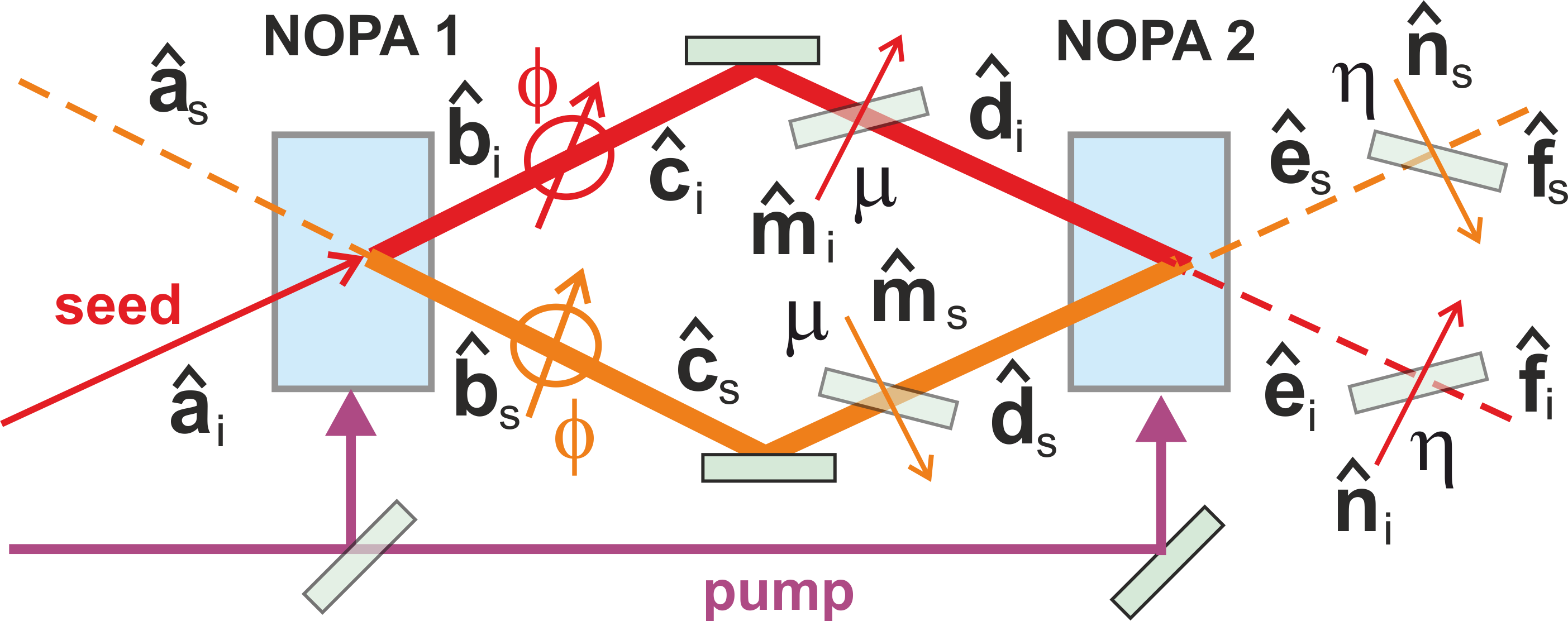}
  \caption{A non-degenerate SU(1,1) interferometer with the notation used in the calculations. The pumping is depicted schematically.}
 \label{fig:SU(1,1)_app}
\end{figure*}

\begin{equation}
  \mean{N_f}\gg1 \,.
\end{equation}
In this case, the discrete value of $N_f$ can be approximated by continuous one, and the phase measurement error can be calculated as
\begin{equation}
  (\Delta\phi)^2
  = \frac{(\Delta N_f)^2 + (\Delta N_d)^2}{\left(\partd{\mean{N_f}}{\phi}\right)^2} \,,
\end{equation}
where [see Eqs.~\eqref{C,S}, \eqref{N}]
\begin{equation}
  \partd{\mean{N_f}}{\phi} = -\mu\eta\sinh2r_1\sinh2r_2\sin2\phi
\end{equation}
where $\Delta N_d\gg1$ is the additional detection error which we take into account in this particular case due to the
relatively small number of photons measured by the
detector. Therefore,
\begin{equation}\label{nonlin_direct_dphi}
  (\Delta\phi)^2
  = \frac{2|C(\phi)|^2|S(\phi)|^2 + A|S(\phi)|^2 + B}{\sinh^22r_1\sinh^22r_2\sin^22\phi},
\end{equation}
where
\begin{equation}\label{AB}
  \begin{array}{rcl}
    A &=& 2\dfrac{1-\mu}{\mu}\sinh^2r_2 + \dfrac{1-\mu\eta}{\mu\eta} \,, \\
    B &=& \dfrac{1-\mu}{\mu}\sinh^2r_2\left(A + \dfrac{1}{\mu}\right)
      + \dfrac{(\Delta N_d)^2}{\mu^2\eta^2}\,.
  \end{array}
\end{equation}

\section{Non-degenerate SU(1,1) interferometer}\label{app:nonlinear2}

We now repeat the calculations of \ref{lin:fields} for the case of a non-degenerate SU(1,1) interferometer (Fig.~\ref{fig:SU(1,1)_app}). The incident fields are   $\hat{{\bf a}}_s$, $\hat{{\bf a}}_i$,
with the subscripts $s$ and $i$ corresponding to the signal and idler modes. The first NOPA transforms them as
\begin{equation}\label{nondeg_in}
  \hat{{\bf b}}_s = \hat{{\bf a}}_s\cosh r_1 + \mathbb{Z}\hat{{\bf a}}_i\sinh r_1 \,,\qquad
  \hat{{\bf b}}_i = \hat{{\bf a}}_i\cosh r_1 + \mathbb{Z}\hat{{\bf a}}_s\sinh r_1 \,,
\end{equation}
\noindent where
\begin{equation}
  \mathbb{Z} = \smatrix{1}{0}{0}{-1} \,.
\end{equation}

\noindent After the phase shift (equal in signal and idler arms), they become
\begin{equation}
  \hat{{\bf c}}_s = \mathbb{O}(\phi)\hat{{\bf b}}_s \,,\qquad
  \hat{{\bf c}}_i = \mathbb{O}(\phi)\hat{{\bf b}}_i \,.
\end{equation}
\noindent The internal losses, also the same in both arms, give
\begin{equation}
  \hat{{\bf d}}_s = \sqrt{\mu}\,\hat{{\bf b}}_s + \sqrt{1-\mu}\,\hat{{\bf m}}_s \,,\qquad
  \hat{{\bf d}}_i = \sqrt{\mu}\,\hat{{\bf b}}_i + \sqrt{1-\mu}\,\hat{{\bf m}}_i \,.
\end{equation}
\noindent After the second NOPA, the quadrature vectors become
\begin{equation}
  \hat{{\bf e}}_s = \hat{{\bf d}}_s\cosh r_2 + \mathbb{Z}\hat{{\bf d}}_i\sinh r_2 \,,\qquad
  \hat{{\bf e}}_i = \hat{{\bf d}}_i\cosh r_2 + \mathbb{Z}\hat{{\bf d}}_s\sinh r_2 \,.
\end{equation}

\noindent Finally, the external losses give
\begin{equation}\label{nondeg_out}
  \hat{{\bf f}}_s = \sqrt{\eta}\,\hat{{\bf e}}_s + \sqrt{1-\eta}\,\hat{{\bf n}}_s \,,\qquad
  \hat{{\bf f}}_i = \sqrt{\eta}\,\hat{{\bf e}}_i + \sqrt{1-\eta}\,\hat{{\bf n}}_i \,.
\end{equation}

\noindent Now introduce the symmetric and antisymmetric modes,
\begin{equation}
  \hat{{\bf a}}_\pm = \frac{\hat{{\bf a}}_s \pm \hat{{\bf a}}_i}{\sqrt{2}} \,, \qquad
  \hat{{\bf b}}_\pm = \frac{\hat{{\bf b}}_s \pm \hat{{\bf b}}_i}{\sqrt{2}} \,,
\end{equation}
and similarly for $\hat{{\bf b}}_{i,s}\,\dots\,\hat{{\bf f}}_{i,s}$, $\hat{{\bf m}}_{i,s}$, $\hat{{\bf n}}_{i,s}$.
Equations for these modes, which can be obtained from (\ref{nondeg_in}-\ref{nondeg_out}), are identical to the equations
for the degenerate case, with the only exception that the squeeze factors for the ``--'' mode are equal to $-r_1$,
$-r_2$. Note that if all incident fields in Eqs.(\ref{nondeg_in}-\ref{nondeg_out}) are  uncorrelated, then the same is
true for the ``$\pm$'' fields. Therefore, the non-degenerate interferometer is equivalent to two independent
degenerate ones.

Note also that in each ``cross-section'' of the original non-degenerate interferometer, the total number of quanta in
the signal and idler beams is equal to the total number of quanta in the corresponding ``cross-section'' of the equivalent
degenerate interferometers, {\it e.g},

\begin{equation}\label{n_b}
  \hat{{\rm b}}_s^\dagger\hat{{\rm b}}_s + \hat{{\rm b}}_i^\dagger\hat{{\rm b}}_i
  = \hat{{\rm b}}_+^\dagger\hat{{\rm b}}_+ + \hat{{\rm b}}_-^\dagger\hat{{\rm b}}_- \,,
\end{equation}
\begin{equation}\label{n_f}
  \hat{{\rm f}}_s^\dagger\hat{{\rm f}}_s + \hat{{\rm f}}_i^\dagger\hat{{\rm f}}_i
  = \hat{{\rm f}}_+^\dagger\hat{{\rm f}}_+ + \hat{{\rm f}}_-^\dagger\hat{{\rm f}}_- \,.
\end{equation}

\section*{References}


%

\end{document}